\documentclass[twocolumn]{aastex631}
\usepackage{lineno, graphicx}
\usepackage{CJK}
\usepackage{hyperref}
\usepackage{soul}
\usepackage{comment}
\usepackage{rotating}

\newcommand{\Hb}{H$\beta$}
\newcommand{\Ha}{H$\alpha$}

\newcommand{\lum}{\ensuremath{\lambda L_{5100}}}
\newcommand{\Mbh}{$M_{\rm BH}$}
\newcommand{\prep}{\texttt{PrepSpec}}
\newcommand{\jav}{\texttt{JAVELIN}}

\newcommand{\q}[1]{\textcolor{red}{#1}}

\begin{document}

\title{The Sloan Digital Sky Survey Reverberation Mapping Project: Insights on Maximizing Efficiency in Lag Measurements and Black-Hole Masses}

\author[0000-0002-0957-7151]{Y. Homayouni}
\affiliation{Department of Astronomy and Astrophysics, The Pennsylvania State University, 525 Davey Laboratory, University Park, PA 16802}
\affiliation{Institute for Gravitation and the Cosmos, The Pennsylvania State University, University Park, PA 16802}

\author[0009-0001-5163-5781]{Yuanzhe Jiang}
\affiliation{Department of Astronomy, School of Physics, Peking University, Beijing 100871, China}
\affiliation{Department of Astronomy, University of Illinois at Urbana-Champaign, Urbana, IL, 61801, USA}

\author[0000-0002-0167-2453]{W. N. Brandt}
\affiliation{Department of Astronomy and Astrophysics, The Pennsylvania State University, 525 Davey Laboratory, University Park, PA 16802}
\affiliation{Institute for Gravitation and the Cosmos, The Pennsylvania State University, University Park, PA 16802}
\affiliation{Department of Physics, 104 Davey Lab, The Pennsylvania State University, University Park, PA 16802, USA}

\author[0000-0001-9920-6057]{C. J. Grier}
\affiliation{Department of Astronomy, University of Wisconsin-Madison, Madison, WI 53706, USA}

\author[0000-0002-1410-0470]{Jonathan R. Trump}
\affiliation{University of Connecticut, Department of Physics, 196 Auditorium Road, Unit 3046, Storrs, CT 06269-3046}

\author[0000-0003-1659-7035]{Yue Shen}
\affiliation{Department of Astronomy, University of Illinois at Urbana-Champaign, Urbana, IL, 61801, USA}
\affiliation{National Center for Supercomputing Applications, University of Illinois at Urbana-Champaign, Urbana, IL, 61801, USA}

\author[0000-0003-1728-0304]{Keith Horne}
\affiliation{SUPA School of Physics and Astronomy, North Haugh, St.~Andrews, KY16~9SS, Scotland, UK}

\author[0000-0002-1763-5825]{Patrick B.\ Hall}
\affiliation{Department of Physics and Astronomy, York University, Toronto, ON M3J 1P3, Canada}

\author[0000-0002-6404-9562]{Scott F. Anderson}
\affiliation{Astronomy Department, University of Washington, Box
351580, Seattle, WA 98195, USA}

\author[0000-0001-6947-5846]{Luis C. Ho}
\affil{Kavli Institute for Astronomy and Astrophysics, Peking University, Beijing 100871, China}
\affil{Department of Astronomy, School of Physics, Peking University, Beijing 100871, China}

\author{D. P. Schneider}
\affiliation{Department of Astronomy and Astrophysics, The Pennsylvania State University, 525 Davey Laboratory, University Park, PA 16802}
\affiliation{Institute for Gravitation and the Cosmos, The Pennsylvania State University, University Park, PA 16802}

\begin{abstract}
Multi-year observations from the Sloan Digital Sky Survey Reverberation Mapping (SDSS-RM) project have significantly increased the number of quasars with reliable reverberation-mapping lag measurements. We statistically analyze target properties, light-curve characteristics, and survey design choices to identify factors crucial for successful and efficient RM surveys. Analyzing 172 high-confidence (``gold") lag measurements from SDSS-RM for the \Hb, \ion{Mg}{2}, and \ion{C}{4} emission lines, we find that the Durbin-Watson statistic (a statistical test for residual correlation) is the most significant predictor of light curves suitable for lag detection. Variability signal-to-noise ratio and emission-line placement on the detector also correlate with successful lag measurements.  We further investigate the impact of observing cadence on survey design by analyzing the effect of reducing observations in the first year of SDSS-RM. Our results demonstrate that a modest reduction in observing cadence to $\sim$1.5 weeks between observations can retain approximately 90\% of the lag measurements compared to twice-weekly observations in the initial year. Provided similar and uniform sampling in subsequent years, this adjustment has a minimal effect on the overall recovery of lags across all emission lines. These results provide valuable inputs for optimizing future RM surveys.
\end{abstract}


%%%%%%%%%%%%%%%%%%%%%%%%%%%%%%%%%%%%%%%%
%%%%%%%%%%%%%%%%%%%%%%%%%%%%%%%%%%%%%%%%
\section{Introduction} \label{sec:intro}
%%%%%%%%%%%%%%%%%%%%%%%%%%%%%%%%%%%%%%%%
%%%%%%%%%%%%%%%%%%%%%%%%%%%%%%%%%%%%%%%%
%Significant advances in observational techniques have played a pivotal role in transforming our view of the universe.
Mounting observations indicate a strong connection between the formation and evolution of galaxies throughout the Universe and the properties of supermassive black holes (SMBHs) residing at their centers. Accurately determining black-hole masses (\Mbh) and characterizing the nature of gas flows in their vicinity are crucial for understanding black-hole -- galaxy coevolution and how black-hole-driven outflows can regulate star formation through feedback mechanisms (e.g., \citealp{Kormendy2013}). However, with the exception of a few SMBHs \citep{gravity18, gravity20, gravity21, gravity24}, spatially resolving the broad-line region (BLR) of active galactic nuclei (AGN) remains a challenge.
%For nearby SMBHs, it is possible to map dynamical effects in the central few parsecs of the SMBH through spatially resolved (interferometric) observations and also obtain a reliable SMBH mass. While spatially-resolved observations have successfully mapped the gas in vicinity of a limited number of nearby SMBHs, resolving the broad-line region (BLR) with this method is currently only feasible for a handful of sources \citep{gravity18, gravity20, gravity21, gravity24}.
Reverberation mapping (RM; \citealt{Blandford1982, Peterson1993}) allows measurement of central black-hole mass and study of the gas dynamics near the black hole by temporally resolving the compact central regions. In its most general form, RM measures the time delay $\tau$ (also known as the lag)  between the continuum and emission-line variability signals in AGN to determine the responsivity-weighted radius of the reprocessing BLR region ($R_{\rm BLR}$). This time delay can be combined with the velocity dispersion of the emission line to estimate the black-hole mass using the virial theorem  (e.g., see the recent review by \citealt{Cackett2021}). \begin{comment}Furthermore, it is also possible to obtain high-fidelity RM observations and use velocity-resolved RM to additionally determine the dynamics of the BLR gas, relative to the central continuum-emitting source through the projection of the BLR into line-of-sight velocity and time delay (e.g., \citealp{Horne2004}).
\end{comment}

Three decades of RM campaigns have compiled a set of $\sim$~100 reliable RM masses in the local universe (e.g., \citealp{Clavel1991, Wanders1997, Kaspi2000, Collier1998, Peterson2004, Bentz2009, Denney2010, Barth2011a, Grier2012, Du2014, Du2016a, Du2016b, Barth2015, Hu2015}). Most of these campaigns used single-object spectrographs, targeting the most variable, local AGNs ($z<$0.3), with relatively low-luminosity (\lum $\,< 10^{45}\, \rm erg\,s^{-1}$) that would have sufficiently short lags to be recovered with a few months of RM monitoring. These campaigns also are predominantly focused on strong emission lines such as the Balmer lines (see \citealt{Bentz2015} for a compilation of \Mbh). However, some campaigns have focused on high-z, luminous quasars using multiple years of RM observations \citep{Kaspi2007, Lira2018}. Some of these campaigns also have  been successful in performing velocity-resolved RM observations \citep{Bentz2008, Bentz2009, Bentz2010b, Grier2013, Bentz2021, Bao2022, U2022, Zastrocky2024}, and also using space-based observations in the UV \citep{Derosa2015, Horne2021, Homayouni23}. 

\begin{comment}These RM measurements have found the existence of a canonical scaling relation between the characteristic radius of the BLR region, $R_{\rm BLR}$, and the AGN continuum luminosity at 5100 $\rm \AA$, also known as the $R-L$ relation \citep{Bentz2013}. \textbf{Although this relation is mostly limited to the sample of low-redshift, low-luminosity, and variable AGNs. The majority of \Mbh\ measurements at higher redshifts are based on simple extrapolation of this local RM sample, which thus enables single-epoch spectroscopy to obtain the radius of BLR and ultimately obtain the SMBH mass. }
However, this relation relies upon limited RM measurements from variability-selected, local AGNs that belong to a narrow range of SMBH mass and accretion rate and are a biased representation of high-redshift ($z>$0.3) quasars (see Figure~1 of \citealt{Shen2015a}). RM observations of quasars that are more representative of high-redshift quasars have revealed deviations from these “single-epoch” black-hole mass estimates, implying that much of our understanding of black-hole growth over cosmic time may be incorrect by up to an order of magnitude (e.g., \citealp{Du2019, FonsecaAlvarez2019}). Furthermore, obtaining reliable RM masses at $1\leq z\leq 2$ is particularly important to study SMBH growth. This redshift window coincides with the peak of SMBH accretion (e.g., \citealp{Brandt2015}), which has been largely missed with RM campaigns on local AGNs.
\end{comment}

Recently, industrial-scale RM programs such as the Sloan Digital Sky Survey Reverberation Mapping (SDSS-RM) project \citep{Shen2015a} and the Australian Dark Energy Survey RM (OzDES) project \citep{King2015} have started to use multi-object spectrographs on survey telescopes to carry out RM observations of hundreds of diverse quasars. %that are more representative of the high-redshift universe.  
%and also campaigns that are dedicated to a specific subclass of AGNS: the Super-Eddington Accreting Massive Black Holes (SEAMBH) Survey \citep{Du2014}. 
While these surveys have been successful, a significant fraction of the sample lacks well-detected lags. This is because extracting reliable and well-defined lag measurements from industrial-scale RM surveys is challenging (see \autoref{sec:timeseries} for more details). This is further complicated because, at higher redshifts, observations are limited to more luminous quasars, which tend to have longer time delays due to the combined effects of cosmological time dilation and the canonical $R-L$ relation \citep{Bentz2013}. Additionally, the inherent lower variability amplitude observed in more luminous quasars (e.g.,\citealp{Vanden2004, MacLeod2010}) further reduces the probability of successful lag recovery.
\begin{comment}Beyond the observational cost of RM studies, industrial-scale RM faces additional challenges due to the increasing volume of light curves compared to dedicated, single-target RM campaigns. In industrial-scale RM studies, often employing multiple computationally intensive time-series analysis methods, is just the first step in lag determination. Subsequent steps are also necessary to identify the most significant and reliable lag measurements. These include applying significance criteria and additional selection cuts to identify spurious light curves and ensure the detection of genuine physical reverberation that would help in identifying the gold sample (see Sections~\ref{sec:alias}, \ref{sec:significance}, and \ref{sec:gold}). These steps cause industrial-scale RM not only observationally expensive but also computationally and data intensive, leading to lengthy analysis times to obtain reliable RM results.  In the era of large surveys like the Rubin Observatory's Legacy Survey of Space and Time (LSST), which will provide access to millions of light curves, developing rapid light curve analysis and diagnostic tools that can efficiently discard light curves unlikely to yield significant and reliable lag measurements would pave the way for faster turnaround times in RM studies.
\end{comment}
Multi-object RM surveys typically target several hundred quasars, such as the 849 in SDSS-RM, 771 in OzDES, and the planned 700 in the 4MOST TiDES program. Industrial-scale RM faces challenges beyond observational costs due to the massive volume of light curves. These challenges primarily stem from the massive volume of data and the complexity of the analysis. Handling hundreds of quasar light curves, each with multiple emission lines, requires specialized software and expertise. Current methods for time-series analysis, outlier rejection, and lag-measurement validation, often involving manual inspection and rigorous statistical tests, are both time-consuming and computationally expensive. While the era of big data has advanced, the field of industrial-scale RM is still relatively young, and the optimization of lag-measurement pipelines is an ongoing challenge. Even with automated methods, human intervention remains essential for quality control and interpretation of results. Visual inspection of hundreds of light curves is a laborious task, requiring significant expertise and time. A typical light curve undergoes multiple stages of processing, including initial vetting, time-series analysis, significance testing, and potential visual inspection or statistical checks. This multi-step approach can extend the analysis timeline for large samples to several months.
%Efficiently analyzing these light curves and selecting the most significant and reliable lag measurements is computationally intensive. 
As future surveys like the Legacy Survey of Space and Time (LSST; e.g., \citealt{Ivezic2019}) produce millions of light curves, developing strategic analysis tools will be crucial for timely and effective RM studies. Although this study focuses on emission-line RM, the methodology developed here can be readily extended to investigate continuum RM, provided that a sufficient number of continuum RM lag measurements are available.

In large-scale surveys like the SDSS-RM, a comprehensive understanding of the interplay between the basic quasar properties, the quasar intrinsic variability characteristics, the survey design, and the observational sensitivity is critical for evaluating and forecasting the program's success in terms of lag-measurement yield and limitations. The detection of reverberation lags relies heavily upon the monitoring program's design, specifically the observing cadence, total observing baseline, presence and distribution of seasonal/weather gaps, and the signal-to-noise ratio (S/N) of the flux measurements.
Previous light-curve simulations of the SDSS-RM survey have already studied the correlation between time-series analysis methods and survey yields \citep{Li2019}. The OzDES campaign has similarly employed light-curve simulations replicating source variability, measurement errors, and observing cadence to assess lag measurement reliability criteria \citep{Penton2022}. However, there has never been a comprehensive assessment connecting physical sample properties and successful lag measurement.
The present work focuses on a sub-sample of the most reliable RM lag measurements from the SDSS-RM survey, specifically those quasars exhibiting well-detected lag measurements (i.e., the ``gold sample", see \autoref{sec:timeseries} for more details). By investigating the physical and statistical properties of this sub-sample, we aim to identify the characteristics of the already \textit{observed} quasars that are most favorable for efficient RM lag measurements from the SDSS-RM survey. We also investigate how reducing the SDSS-RM cadence affects the number of gold-lag measurements. This analysis will inform future RM surveys, such as the 4MOST TiDES program  \citep{deJong2019, Swann2019}, by demonstrating the potential for achieving comparable lag results with less-frequent observations.%, which aims to conduct RM observations of 700 quasars over five years, continuing to monitor some of the previously studied RM fields. 

In \autoref{sec:data} we give an overview of the SDSS-RM survey, sample selection, data, and data processing. \autoref{sec:timeseries} describes our custom time-series analysis pipeline and strategies for reliable lag identification, and the sample used for this work. We present the analysis of physical AGN properties in \autoref{sec:target_phys}, and discuss connections to statistical light-curve properties in \autoref{sec:light_curve}. In \autoref{sec:cadence} we describe the impact of reduced cadence on lag success and recovery. We discuss the implications of our assessments in \autoref{sec:discussion}. Throughout this work, we adopt a $\Lambda$CDM cosmology with $\Omega_{\Lambda}$ = 0.7, $\Omega_M$ = 0.3, and $H_0$ = 70 km $\mathrm{s^{-1}\,Mpc^{-1}}$.
%%%%%%%%%%%%%%%%%%%%%%%%%%%%%%%%%%%%%%%%%%%%%
%%%%%%%%%%%%%%%%%%%%%%%%%%%%%%%%%%%%%%%%%%%%%
\section{Data} \label{sec:data}
%%%%%%%%%%%%%%%%%%%%%%%%%%%%%%%%%%%%%%%%%%%%%
%%%%%%%%%%%%%%%%%%%%%%%%%%%%%%%%%%%%%%%%%%%%%

%%%%%%%%%%%%%%%%%%%%%%%%%%%%%%%%%%%%%%%%%%%%%
\subsection{SDSS-RM Survey Overview} \label{sec:sdss-rm}
%%%%%%%%%%%%%%%%%%%%%%%%%%%%%%%%%%%%%%%%%%%%%
The SDSS-RM project is a time-domain multi-object spectroscopic RM (MOS-RM) survey that simultaneously monitored 849 broad-line quasars at $0.1<z<4.5$ in a single 7 $\rm deg^2$ field \citep{Shen2015a}. Due to its simple magnitude-limited selection criteria of $i_{\rm psf} \leq 21.7$, the SDSS-RM survey has significantly expanded the parameter space of AGNs studied with RM, which allows for the investigation of luminous quasars beyond the local universe. An overview of the SDSS-RM design, observing strategy, and target selection is reported in \citet{Shen2015a}, with details of sample properties reported in \citet{Shen2019a}. The primary goal of SDSS-RM is to provide SMBH mass measurements for a broad range of redshifts and luminosities \citep{Shen2016a, Grier2017, Grier2019, Homayouni2020, Shen2024}. However, it also has been successful in enabling many ancillary studies of quasar variability and other physical properties (e.g., \citealp{Shen2015b, Sun2015, Denney2016b, Denney2016a, Grier2016, Dexter2019, Hemler2019, Homayouni2019,Wang2019, Li2019, FonsecaAlvarez2019, DallaBonta2020, Li2021, Li2023, Fries2023}).

%%%%%%%%%%%%%%%%%%%%%%%%%%%%%%%%%%%%%%%%%%%%%
\subsection{Data Processing} 
%%%%%%%%%%%%%%%%%%%%%%%%%%%%%%%%%%%%%%%%%%%%%

SDSS-RM observed every year during 2014--2020 as part of SDSS-III \citep{Eisenstein2011} and SDSS-IV \citep{Blanton2017}, and observations of a subset of the SDSS-RM quasar field are continuing as part of the SDSS-V Black Hole Mapper (BHM) program. The spectroscopic monitoring required for SDSS-RM was provided by the BOSS spectrograph \citep{Smee2013} on the SDSS telescope \citep{Gunn2006}. The observations achieved an average cadence of 4 days with 32 epochs in the first year, $\sim$ 12 epochs per year between 2015 -- 2017, and $\approx$ 6 epochs per year (monthly cadence) during 2018 -- 2020, totalling 90 spectroscopic epochs over the course of seven years of monitoring. SDSS-RM was also accompanied by optical photometric monitoring in the $g$ and $i$-band from the 2.3 m Bok telescope at Steward Observatory and the 3.6 m Canada France Hawaii Telescope (CFHT) to enhance the continuum light curve to facilitate RM measurements. These observations largely overlap with the spectroscopic observation window, and have similarly higher cadence in 2014, and reduced cadence in the subsequent years. \citet{Kinemuchi2020} describe the photometric component of the SDSS-RM program, and the associated data reduction. Furthermore, the SDSS-RM field overlaps with the PanSTARRS-1 MD07 Medium Deep Field \citep{Kaiser2010}, providing additional multi-band photometry (2010~--~2013). We also incorporate Zwicky Transient Facility data (2018~--~2020, \citealp{Bellm2019}), which together with Pan-STARRS data, extend the light curve baseline to 11 years (2010~--~2020) for investigation of longer lags in SDSS-RM observations.

The spectroscopic data are initially processed using the standard SDSS pipeline, followed by a custom calibration pipeline to improve flux calibration \citep{Shen2015a}. The data are then further reprocessed using the  \texttt{PrepSpec} software \citep{Shen2015a, Shen2016a} to improve relative flux calibration assuming the flux of narrow emission lines do not intrinsically vary throughout the RM campaign. \prep\ then models the spectra, continuum, and the broad line and produces light curves. \prep\ also produces measurements of mean and root mean square (rms) residual line profiles, line widths, and light curves for each of the model components. It also returns several other statistical products (see Section \ref{sec:light_curve} for more details).

The photometric data from different instruments, facilities, and filters are combined to account for observatory seeing variations, calibration issues for each filter response, telescope throughput, and other site-dependent effects. We adopt the publicly available \texttt{PyCali} code \citep{Li2014} to perform the light curve merging process. \texttt{PyCali} uses a Bayesian framework to achieve this, allowing for the adjustment of individual light curve flux uncertainties. This process mitigates overestimation and underestimation of the uncertainties reported in the light curves. We normalize all light curves to the flux of synthetic photometry in the $r$-band, where we measure the synthetic photometry by convolving the \prep-corrected spectra with SDSS filters responses \citep{Fukugita1996} to establish a common reference flux level. 

%%%%%%%%%%%%%%%%%%%%%%%%%%%%%%%%%%%%%%%%%%%%%
\subsection{Sample Selection} 
%%%%%%%%%%%%%%%%%%%%%%%%%%%%%%%%%%%%%%%%%%%%%
In less than one decade, the SDSS-RM program has substantially expanded the set of reliable SMBH masses measured through RM to $\sim$300 quasars out to $z>$3 \citep{Shen2016a, Grier2017, Grier2019, Shen2019b, Homayouni2020, Shen2024}. %While the first-season results from 6-month SDSS-RM observations were reported by \citet{Shen2016a, Grier2017} for shorter lags ($\tau< 100$~days), the lags reported by \citet{Shen2016a} are only based on spectroscopic observations and lack the accompanying SDSS-RM photometry to improve the continuum light curve for RM lag measurements. 
\citet{Grier2017} measured \Hb\ and \Ha\ lags from the first-season observations from 2014. For longer lags in \Ha, \Hb, \ion{Mg}{2}, and \ion{C}{4}, multi-year RM lag results were reported by \citet{Grier2019}, \citet{Shen2019b}, \citet{Homayouni2020}, and \citet{Shen2024}. In this work, we focus on the targets with the highest-quality lag measurements from these previous SDSS-RM studies. Our subsample is drawn from those SDSS-RM studies that use the improved continuum light curves from ground-based photometry. We select lag measurements that have quality ratings of 4 or 5 from \citet{Grier2017, Grier2019, Shen2024} (also known as the gold sample) and the most reliable lag measurements with individual false-positive rate (FPR) of $<10\%$ as defined in \citet{Homayouni2020}. In total, we have 137 RM lag measurements that are flagged as the ``gold sample" with 26 lag measurements using \Hb\ \citep{Grier2017}, 24 from \ion{Mg}{2} \citep{Homayouni2020}, and 16 in \ion{C}{4} \citep{Grier2019} reported from the early-year SDSS-RM studies, and 37 in \Hb, 32 in \ion{Mg}{2}, and 37 in \ion{C}{4} from a recent SDSS-RM investigation \citep{Shen2024}. In this work, the one-year results from \citet{Grier2017}, and four-year results from \citet{Grier2019} and \citet{Homayouni2020} are referred to as the ``early-year" SDSS-RM lag results. We refer to the most recent lag measurements from \citet{Shen2024} combining 7 years of spectroscopy with 11 years of photometry as the ``7-year" SDSS-RM results. We also refer to the larger SDSS-RM sample of 849 quasars as the ``parent sample".
%%%%%%%%%%%%%%%%%%%%%%%%%%%%%%%%%%%%%%%%%%%%%
%%%%%%%%%%%%%%%%%%%%%%%%%%%%%%%%%%%%%%%%%%%%%
\section{Time-Series Analysis } \label{sec:timeseries}
%%%%%%%%%%%%%%%%%%%%%%%%%%%%%%%%%%%%%%%%%%%%%
%%%%%%%%%%%%%%%%%%%%%%%%%%%%%%%%%%%%%%%%%%%%%

Ideally, RM observations rely on densely-sampled data to precisely track AGN variability and perform robust time-series analysis. However, observing limitations such as weather loss and telescope-scheduling constraints can lead to sparsely sampled light curves. This sparsity in a light curve requires interpolation between observing epochs to accurately measure lags and their associated uncertainties. Variability tracking is further challenged in multi-year survey observations because of the presence of seasonal gaps in the data. Historically, RM observations use the Interpolated Cross Correlation Function (ICCF) method to perform time-series analysis  \citep{Gaskell1986, Gaskell1987, Peterson2004}. To perform time-series analysis for SDSS-RM light curves, we adopt approaches that are flexible in handling multi-year observations that are associated with observing gaps: \texttt{JAVELIN} \citep{Zu2011}, \texttt{CREAM} \citep{Starkey2016}, and more recently \texttt{PyROA} \citep{Donnan2021}. These methods model the light curve behavior during observing gaps assuming stochastic variability of quasar light curves, whereas the  ICCF method relies on linear interpolation between epochs. Comparison between these methods reveals that the ICCF technique results in higher uncertainty in lag measurements compared to the other model-dependent methods, particularly in industrial-scale datasets like the SDSS-RM program and similar reverberation-mapping campaigns \citep{Li2019, Yu2020}. While established methods like \jav\ and ICCF have been well-studied for reverberation mapping, more recent techniques like  \texttt{PyROA} lack comprehensive comparisons with these existing methods. 
%\yh{Future investigation could involve a thorough evaluation of \texttt{PyROA} alongside \jav\ and ICCF to assess their relative strengths and weaknesses in the context of reverberation-mapping analysis, particularly for datasets like those from the SDSS-RM program.}

%%%%%%%%%%%%%%%%%%%%%%%%%%%%%%%%%%%%%%%%%%%%%
\subsection{Alias Removal}\label{sec:alias}
%%%%%%%%%%%%%%%%%%%%%%%%%%%%%%%%%%%%%%%%%%%%%
Sparsely-sampled data in RM time-series analysis can lead to artifacts in lag-measurement results. Specifically, posterior lag distributions obtained from methods like \jav\ and \texttt{CREAM}, and the cross-correlation centroid distribution (CCCD) from ICCF, may generate a pronounced primary peak that is accompanied by the presence of secondary peaks, generally less significant than the primary. These secondary peaks are likely aliases, arising from the interplay of low-cadence sampling and noisy measurements, where the Markov chain Monte Carlo (MCMC) algorithm, in attempting to fit the light curve, may introduce spurious correlations in weakly variable segments, leading to these secondary peaks. The presence of multiple spurious peaks, or aliases, within the lag posterior distribution function (PDF) can significantly compromise lag estimation. While one alias may appear dominant, the presence of others indicates a poorly constrained lag measurement. This skews the PDF, potentially leading to biased lag estimates and inflated uncertainties. The presence of seasonal gaps further complicates aliasing in RM analysis. Methods like \jav\ model the light curve throughout the entire series, including seasonal gaps, which can lead to spurious lag solutions that coincide with the gaps. This occurs because the algorithms may misinterpret the absence of data during seasonal gaps as a correlated signal, leading to artificial peaks in the lag posterior distribution. Therefore, to mitigate the effects of aliasing in the lag PDF, a systematic approach to removing these artifacts is necessary.

Prior studies within the SDSS-RM collaboration have employed a consistent alias removal approach. This method incorporates a weighting function on the lag PDF that assigns lower weights to lag values corresponding to segments with minimal overlap between the continuum and line light curves \citep{Grier2017, Grier2019, Homayouni2020, Shen2024}. Essentially, lags exhibiting a complete absence of overlap in the light curves at a specific time delay ($\tau$) are deemed less probable by the weighting scheme. The weighting scheme incorporates two key factors to address aliasing: the number of overlapping data points and the influence of continuum variability; the final weighting scheme is the convolution between these two terms. The first factor is captured by $[N(\tau)/N(0)]^2$, where $N(\tau)$ represents the number of overlapping data points at a specific time lag and $N(0)$ denotes the number of overlapping points at zero lag $\tau = 0$. Lower values of this term indicate minimal overlap. The second factor is introduced by the continuum auto-correlation function (ACF). A narrow ACF signifies a rapidly varying continuum, making it challenging to distinguish the light curve's behavior within the gaps. Conversely, a broad ACF indicates a slowly varying continuum, where gaps have a minimal impact on the measured lag. 

%%%%%%%%%%%%%%%%%%%%%%%%%%%%%%%%%%%%%%%%%%%%%
\subsection{Significance Criteria}\label{sec:significance}
%%%%%%%%%%%%%%%%%%%%%%%%%%%%%%%%%%%%%%%%%%%%%
While the alias-removal approach eliminates spurious peaks and aliases from the lag PDF, further assessment is necessary to identify the most reliable lag measurements. This becomes important since the alias removal process might inadvertently suppress genuine peaks alongside aliases, potentially leaving the alias-rejected lag PDF with a weak primary peak that is difficult to confidently identify. Furthermore, the lag PDF may identify lags that are statistically consistent with zero, offering no meaningful physical interpretation. To address these concerns, we employ additional selection criteria to ensure the final reported lags are statistically robust and correspond to a true physical reverberation process.

To identify statistically significant lag measurements, we employ a set of selection criteria established through a combined approach of statistical assessment and visual inspection of the lag PDF and the light curves. Ideally, these criteria are chosen to achieve a false-positive rate of $\approx$ 10\%. This target rate is informed by simulations from \cite{Shen2015a} and considers the data quality within the SDSS-RM program, including factors such as light-curve cadence, signal-to-noise ratio (S$/$N), and the presence of seasonal gaps. Previous SDSS-RM lag measurements have employed a suite of criteria to assess the significance of lag detections, which include:
\begin{itemize}
    \item $f_{\rm peak}$: The fraction of the weighted lag posterior integrated within the primary peak can be used as a criterion for significance. This metric assesses whether the primary peak concentrates a sufficient fraction of the posterior probability to be considered a reliable lag measurement.
    \item $r_{\rm max}$: The maximum Pearson cross-correlation coefficient $r_{\rm max}$ between the continuum and the line light curves can serve as an indicator of correlated variability and physical time lag due to reverberation processes.
    \item Lag $\rm |S/N|$: We can assess the consistency of a lag measurement with zero lag by considering the absolute value of the measured lag and its associated uncertainty, and can identify if a lag measurement is consistent with zero.
    \item rms variability S/N: This criterion effectively removes cases where the light curves exhibit minimal intrinsic variability, thereby preventing the lag detection methods from erroneously identifying monotonic trends or spurious correlations between noisy light curves. %Such light curves can mislead lag detection methods, causing them to spuriously associate with monotonic trends or noisy correlations within the light curve. %which can result in the lag detection methods latching onto monotonic trends or spurious correlations between noisy light curves.
\end{itemize}

These are our general selection criteria, though some studies have adopted more stringent requirements \citep{Homayouni2019}.

%%%%%%%%%%%%%%%%%%%%%%%%%%%%%%%%%%%%%%%%%%%%%
\subsection{Selection of the ``Gold Sample"}\label{sec:gold}
%%%%%%%%%%%%%%%%%%%%%%%%%%%%%%%%%%%%%%%%%%%%%
\begin{figure*}[t]
\centering
\includegraphics[width=\textwidth]{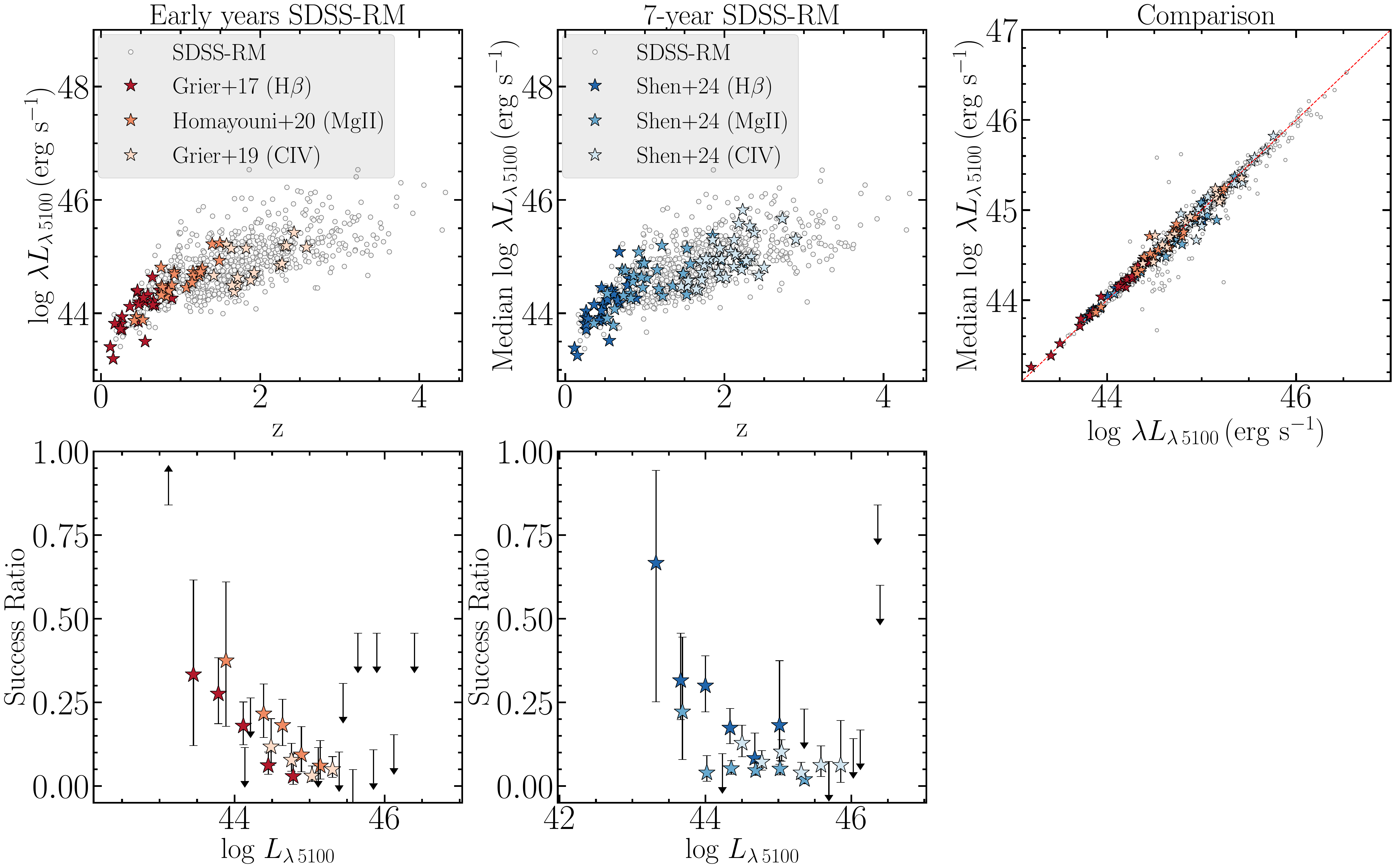}
 \caption{Comparison of the host subtracted \lum\ continuum luminosity in the gold sample of SDSS-RM quasars. The SDSS-RM parent sample is illustrated in gray, open symbols. The top left panel illustrates the gold lag measurements (colored symbols) from the first-year and four-year (early results) SDSS-RM measurements in \citet{Grier2017, Homayouni2020, Grier2019}. The top middle panel shows the 7-year lag measurements in \Hb, \ion{Mg}{2}, and \ion{C}{4} from \citet{Shen2024}. The continuum luminosity here are host-subtracted using the principle component analysis as reported by \citet{Shen2015b}. Overall, there is no notable difference between the luminosity in the SDSS-RM parent sample and the targets with the most reliable lag measurements in either the early SDSS-RM or 7-year lag measurements (see Table~\ref{tab:table2} for a comparison of median values in each work). The top right panel compares the first-year with the 7-year continuum luminosity in the SDSS-RM sample. Overall, the targets do not show a significant change in their luminosity. The bottom panels show the gold-lag success fraction as a function of the early-year (left) and 7-year (right) continuum luminosity. The lag success ratio exhibits a decreasing trend with increasing source luminosity. This could be due to the longer time delays of more luminous targets (e.g., \citealt{Bentz2013}), which may require longer observational baselines not covered by the studies in the current work. Here and in subsequent figures, the 68\% uncertainty in the success ratio has been calculated using small-number statistics \citep{Gehrels1986}. 
  }
\label{fig:luminosity}
\end{figure*}

\begin{figure*}
\centering
\includegraphics[width=0.86\textwidth]{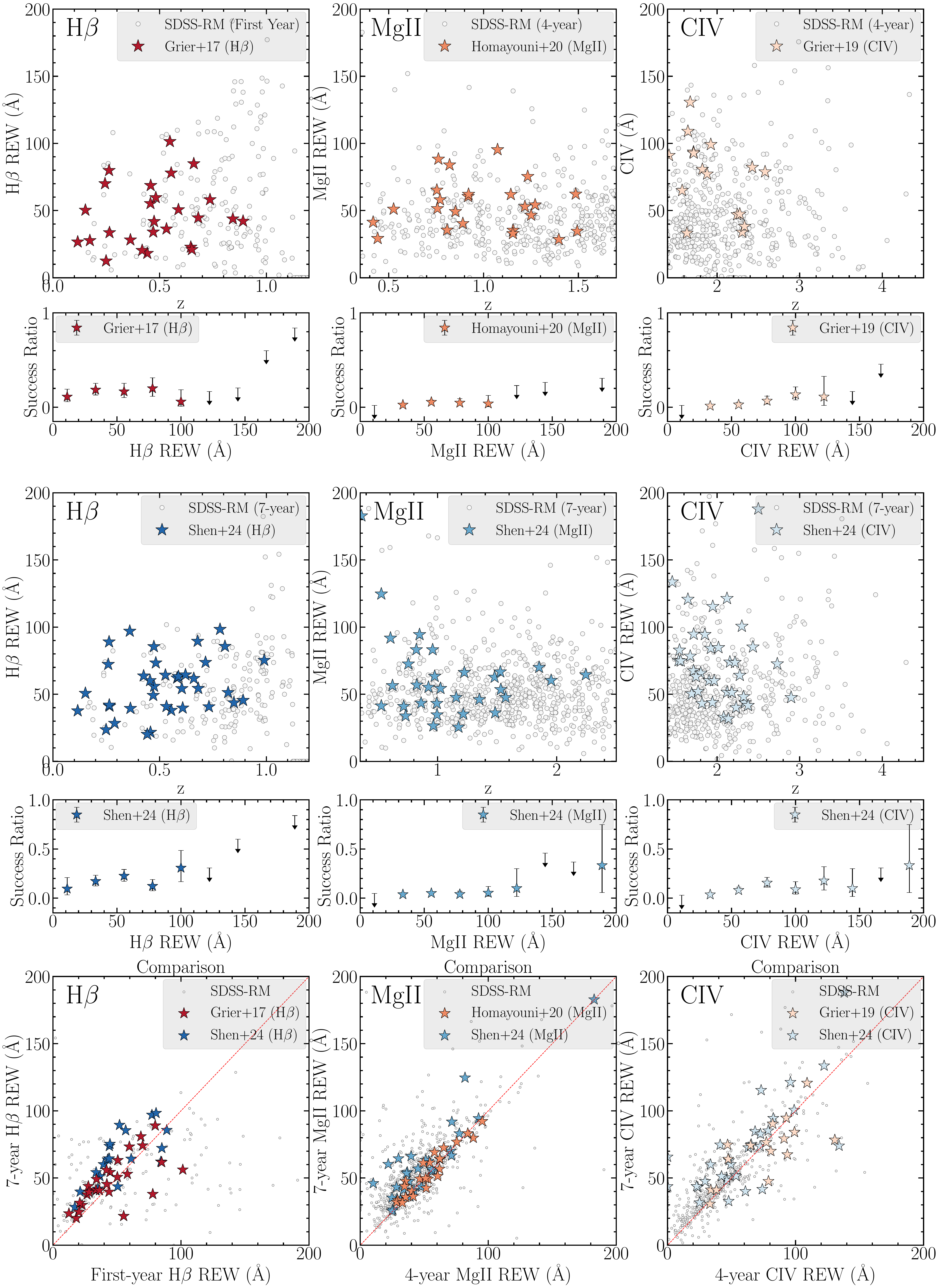}
 \caption{Comparison of the SDSS-RM REW for \Hb\ (left column), \ion{Mg}{2} (middle column), and \ion{C}{4}\ (right column). The gold sample in each study is identified by the colored stars, where early SDSS-RM results are shown in a reddish color palette and the 7-year results are shown in blueish tones and the parent sample in each case is shown with gray symbols, which reveals a uniform distribution when compared to the parent population. For each work, we also include a panel that illustrates the gold-lag success fractions as a function of REW, which reveals that there is no discernible trend evident for the success ratio and larger REW. The panels in the bottom row show the REW comparison between the early-year and the 7-year lag results. Table~\ref{tab:table2} demonstrates that gold lags in the \ion{C}{4} sample have marginally larger REWs compared to the other emission lines. While these larger REWs may be expected to influence the gold-lag success ratio, the observed correlation is relatively weak.}
\label{fig:ew}
\end{figure*}
Large-scale RM surveys like the SDSS-RM program are inherently susceptible to a certain level of false-positive lag detections. Limited sampling cadence and the presence of seasonal gaps can lead to lag PDFs exhibiting well-defined peaks that satisfy our significance criteria. These peaks may arise from noisy light curves or spurious correlations within the light curve, rather than reflecting a true physical reverberation process.
Therefore, \citet{Grier2017, Homayouni2020, Grier2019} and \citet{Shen2024} have employed a secondary selection process to identify the most reliable lag measurements, often referred to as the gold sample. This selection process has primarily relied on visual inspection of the light curves and lag PDFs to identify these high-confidence lag measurements. Distinguishing genuine reverberation from spurious detections remains a challenge, with the specific criteria varying between studies. Key factors for gold-lag measurement include a unimodal lag probability density function, indicating a single dominant lag value, and sufficient light curve variability. Furthermore, consistency across different analysis methods and a good fit of the applied model to the light curve data provide strong supporting evidence for the validity of the measured lag. Other approaches such as simulation-based methods and quantitative measures can provide objective assessments of lag reliability, but they often require significant computational resources, as demonstrated by \cite{Homayouni2020} and \cite{Yu2021}. This is especially true for identifying a gold sample in industrial RM studies, which involves time-intensive simulations of lightcurves.
\begin{comment}Some studies have also explored quantitative measures that would provide an objective approach, but can be computationally expensive as demonstrated by \cite{Homayouni2020}. Similarly, simulation-based methods, such as those employed by \cite{Yu2021}, can also be computationally expensive. In either case, identification of a gold sample in industrial RM studies presents a significant challenge due to the time-intensive nature of simulating lightcurves. 
\end{comment}

\startlongtable
\begin{deluxetable*}{ccccc}
%\tabletypesize{\scriptsize}
\tablecaption{Gold Lag Summary in SDSS-RM Survey \label{tab:table1}}
% \left\langle F \right\rangle
\tablehead{
\colhead{Targeted Emission Line} & \colhead{Results} & \colhead{$\rm N_{Sample\,Total}$} & \colhead{$\rm N_{Gold}$} & \colhead{Gold-Lag Success Ratio} \\
%\colhead{} & \colhead{} & \colhead{} & \colhead{} & \colhead{log($\rm erg\,s^{-1}$)} & \colhead{$\rm\AA$} & \colhead{} & \colhead{} & \colhead{} & \colhead{} \\
}
\startdata
\Hb\ (first-year) & \citet{Grier2017}& 222 & 26 &  12\% \\ 
MgII (four-year) & \citet{Homayouni2020} & 193 & 24 &  12\% \\ 
CIV (four-year) & \citet{Grier2019} & 349 & 16 &  5\% \\
\Hb\ (seven-year) & \citet{Shen2024} & 186 & 37 &  20\% \\
MgII (seven-year) & \citet{Shen2024} & 714 & 32 &  4\% \\
CIV (seven-year) & \citet{Shen2024} & 494 & 37 &  7\%\\
\enddata
% \tablecomments{
%\footnotesize
%Overall 137 RM lag measurements are identified as gold-lag measurements. The 7-year study of \citet{Shen2024} combined 7-years of spectroscopy with 11-years of photometry to obtain the RM lag measurements.}
\end{deluxetable*}

From the parent sample of 849 quasars in the SDSS-RM sample, the applicable redshift range for \Hb\ lags ($0.3< z < 1.14$) has limited the usable sample to 222 objects. From this subsample, \citet{Grier2017} achieved successful lag measurements for 44 quasars, with 26 classified as gold. Similarly, RM measurements of \ion{Mg}{2} in \citet{Homayouni2020} were restricted to $0.3<z<1.7$, leading to 193 quasars in the \ion{Mg}{2}-subsample, which yielded successful lag measurements in 57 quasars, out of which 24 were identified as gold. \citet{Grier2019} investigated \ion{C}{4} lags, but their analysis was limited to $1.4<z<4.5$ due to the upper redshift limit of the SDSS-RM parent sample. Out of 349 quasars, successful lag measurements were obtained for 48, only 16 of which were classified as gold. The resource-intensive nature of RM, coupled with the low yield of gold-lag measurements, motivates further investigation in the current work. Similarly for the 7-year lag results, \citet{Shen2024} recently analyzed a sample of 187 quasars for \Hb, 714 for \ion{Mg}{2}, and 494 for \ion{C}{4}, identifying 37, 32, and 37 gold measurements, respectively. Table~\ref{tab:table1} provides a brief summary of the overall success fraction of gold-lag measurements in SDSS-RM.

%%%%%%%%%%%%%%%%%%%%%%%%%%%%%%%%%%%%%%%%%%%%%
%%%%%%%%%%%%%%%%%%%%%%%%%%%%%%%%%%%%%%%%%%%%%
\section{Target Properties} \label{sec:target_phys}
%%%%%%%%%%%%%%%%%%%%%%%%%%%%%%%%%%%%%%%%%%%%%
%%%%%%%%%%%%%%%%%%%%%%%%%%%%%%%%%%%%%%%%%%%%%

We examine the extent to which inherent target properties, including both physical characteristics and statistical light-curve signatures, produce the most reliable lag measurements. To ensure that target properties do not vary significantly relative to the baseline observations, we study the properties within each study. These included one-year and four-year RM lag measurements reported by \citet{Grier2017, Homayouni2020}, and \citet{Grier2019} alongside the recent seven-year results by \citet{Shen2024}. This comparison aims to mitigate any significant discrepancies potentially arising from the difference in the baseline of observations (short vs. long). For each property, we also investigate the correlation between gold-lag success ratio and the property's value. A comparative analysis of the median values for the parent and gold samples is also provided in Table~\ref{tab:table2} (physical target properties) and Table~\ref{tab:table3} (statistical light-curve properties).

\begin{table*}[!t]
\caption{Comparative Analysis of Sample Properties: Gold and Parent \label{tab:table2}}
\begin{center}
\begin{tabular}{c|cc|cc}
\hline
\hline
Emission Line & \multicolumn{2}{c}{\lum} & \multicolumn{2}{c}{REW} \\
{} & \multicolumn{2}{c}{log($\rm erg\,s^{-1}$)} & \multicolumn{2}{c}{\AA} \\
{} & Parent & Gold & Parent & Gold \\
\hline
\hline
\Hb\ (first-year) & $44.37^{+0.37}_{-0.43}$ & $44.14^{+0.24}_{-0.42}$ & $44.8^{+67.0}_{-30.3}$ & $42.9^{+27.1}_{-20.2}$ \\ [5pt]
MgII (four-year) & $44.73^{+0.5}_{-0.41}$ & $44.6^{+0.2}_{-0.29}$ & $44.7^{+26.0}_{-14.9}$ & $52.2^{+16.3}_{-17.1}$ \\ [5pt]
CIV (four-year) & $45.09^{+0.4}_{-0.41}$ & $44.86^{+0.33}_{-0.32}$ & $48.0^{+43.5}_{-23.7}$ & $79.8^{+17.2}_{-38.1}$ \\ [5pt]
\Hb\ (seven-year) & $44.32^{+0.42}_{-0.43}$ & $44.15^{+0.31}_{-0.33}$ & $54.1^{+26.8}_{-24.9}$ & $54.4^{+23.4}_{-15.2}$ \\ [5pt]
MgII (seven-year) & $44.75^{+0.49}_{-0.42}$ & $44.65^{+0.24}_{-0.35}$ & $49.3^{+23.9}_{-17.6}$ & $55.8^{+27.3}_{-19.7}$ \\ [5pt]
CIV (seven-year) & $45.03^{+0.39}_{-0.42}$ & $44.93^{+0.37}_{-0.34}$ & $47.7^{+37.3}_{-20.0}$ & $67.5^{+28.7}_{-23.7}$ \\ [5pt]
\hline
\hline
\end{tabular}
\tablecomments{\footnotesize
The reported uncertainty on each quantity is the 16th and 84th percentile values for each population.} 
\end{center}

\end{table*}

%\Hb\ (first-year) & 44.4 $\pm$ 0.3 & 44.1 $\pm$ 0.4 & 44.8 $\pm$ 43.2 & 42.9 $\pm$ 23.4 \\ 
%MgII (four-year) & 44.7 $\pm$0.5 & 44.6 $\pm$ 0.2 & 44.7 $\pm$ 17.5 & 52.2 $\pm$ 16.9 \\
%CIV (four-year) & 45.1 $\pm$0.4 & 44.9 $\pm$ 0.4 & 48.0 $\pm$ 30.4 & 79.8 $\pm$ 25.4 \\
%\Hb\ (seven-year) & 44.3 $\pm$0.4 & 44.2 $\pm$ 0.5 & 54.1 $\pm$ 27.9 & 54.4 $\pm$ 20.3 \\
%MgII (seven-year) & 44.7 $\pm$0.5 & 44.7 $\pm$ 0.4 & 49.3 $\pm$ 20.0 & 55.8 $\pm$ 19.9 \\
%CIV (seven-year) & 45.0 $\pm$0.4 & 44.9 $\pm$ 0.3 & 47.7 $\pm$ 26.3 & 67.5 $\pm$ 25.5 \\

\subsection{Continuum Luminosity}\label{sec:luminosity}
Intrinsic AGN luminosity is considered the primary driver of reverberation lag times. More luminous quasars tend to have larger time lags as observed by the $R-L$ relation \citep{Bentz2013, FonsecaAlvarez2019}, which matches basic photoionization expectations ($R_{\rm BLR} \propto \sqrt{L}$). Furthermore, luminous quasars often exhibit lower amplitude variability \citep{MacLeod2012, Vanden2004}, potentially reducing the likelihood of detecting a reliable lag measurement within short observing campaigns. To explore this connection, we adopt the \lum\ luminosity as a measure of the quasar continuum luminosity. Whenever possible, we utilize directly measured \lum\ values. For sources lacking direct \lum\ measurements, we reconstruct them using $\lambda\,L_{3000}$, $\lambda\,L_{1350}$, and the bolometric corrections from \citet{Richards2006}.
Figure~\ref{fig:luminosity} illustrates the redshift and luminosity distribution of our targets. We performed a global comparison of the continuum luminosity within the gold samples of early SDSS-RM studies and the current 7-year lag measurements \citep{Grier2017, Grier2019, Homayouni2020, Shen2024}. This analysis did not reveal any statistically significant changes in the continuum luminosity distribution over the multi-year analysis. Additionally, we find no dependence between lag measurement success and source luminosity. The median continuum luminosity of the gold sample (in logarithmic scale) is $44.1^{+0.2}_{-0.4}$, $44.6^{+0.2}_{-0.3}$, $44.9^{+0.3}_{-0.3}$ for \Hb, \ion{Mg}{2}, and \ion{C}{4}, respectively, for the early SDSS-RM lag measurements, which is similar to the median range of $44.4^{+0.4}_{-0.4}$, $44.7^{+0.5}_{-0.4}$, $45.1^{+0.4}_{-0.4}$ $\rm erg \,s^{-1}$ for the parent sample of each targeted emission line. Here, the reported uncertainty is computed from the 16th and 84th percentiles.
%$44.1\pm0.1$ $\rm erg \,s^{-1}$, $44.6\pm0.1$ $\rm erg \,s^{-1}$, and $44.9 \pm0.2$ $\rm erg \,s^{-1}$ for \Hb, \ion{Mg}{2}, and \ion{C}{4} respectively for the early SDSS-RM lag measurements, which is similar to the median range of $44.4 \pm0.2$, $44.7 \pm 0.05$, and $45.1\pm0.1$ $\rm erg \,s^{-1}$ in the parent sample of each targeted emission line. Here the uncertaintly on the median vales are computed through bootstrap method. 
We find similar ranges for the gold-lag measurements in the 7-year results of \citet{Shen2024}.

\subsection{Rest-frame Equivalent Width}
The rest-frame equivalent width (REW) of emission lines is another potential factor influencing successful lag measurement success. We investigate whether targets with larger REWs correlate with the most reliable lag measurements. For each emission line, we examine the potential relationship between its REW using PrepSpec outputs. REWs are measured from the continuum and flux provided by PrepSpec, utilizing either one-year, four-year, or seven-year average spectra. Figure~\ref{fig:ew} shows the comparisons between early-year and 7-year SDSS-RM data for the same emission lines, which reveal relatively consistent REWs over time. Furthermore, we find no direct correlation between the fraction of quasars with gold-lag measurements. Figure~\ref{fig:ew} depicts how the gold samples in each study are uniformly distributed within their respective parent samples, and the gold-lag success fraction does not  reveal any specific trends with REW. While the one-year and four-year PrepSpec spectral fits did not incorporate any \ion{Fe}{2} emission lines, the seven-year data included an \ion{Fe}{2} template. Consequently, the 7-year measurements of \citealt{Shen2024} (blue data points in Figure~\ref{fig:ew}) appear to be systematically offset above the 1:1 line, suggesting that for all the three emission lines, the gold-lag estimates may be biased towards larger REWs. Comparison of the median REW values between the gold and parent samples in Table~\ref{tab:table2} indicates a higher success rate for the gold-lag sample in \ion{C}{4} and marginal for \ion{Mg}{2}.%The median of the REWs in \Hb\ gold-lag measurements are 43 $\pm$ 6 $\rm km\,\,s^{-1}$, and the REWs in the \Hb\ parent sample is 45 $\pm$ 4 $\rm km\,\,s^{-1}$. Similarly, we find a median of 52 $\pm$ 4.5 $\rm km\,\,s^{-1}$ in \ion{Mg}{2} gold lag sample, while the parent sample has a REW of 41.4. \q{For \ion{C}{4}, the gold lags have a mean REW of 80 $\pm$ 10$\rm km\,\,s^{-1}$ whereas the parent sample REW is 42 $\rm km\,\,s^{-1}$}. 
%%%%%%%%%%%%%%%%%%%%%%%%%%%%%%%%%%%%%%%%%%%%%%%%%%%%%%%
%%%%%%%%%%%%%%%%%%%%%%%%%%%%%%%%%%%%%%%%%%%%%%%%%%%%%%%
\subsection{Redshift and Observed-Frame Wavelength}
%%%%%%%%%%%%%%%%%%%%%%%%%%%%%%%%%%%%%%%%%%%%%%%%%%%%%%%
%%%%%%%%%%%%%%%%%%%%%%%%%%%%%%%%%%%%%%%%%%%%%%%%%%%%%%%

\begin{figure*}
\centering
\includegraphics[width=\textwidth]{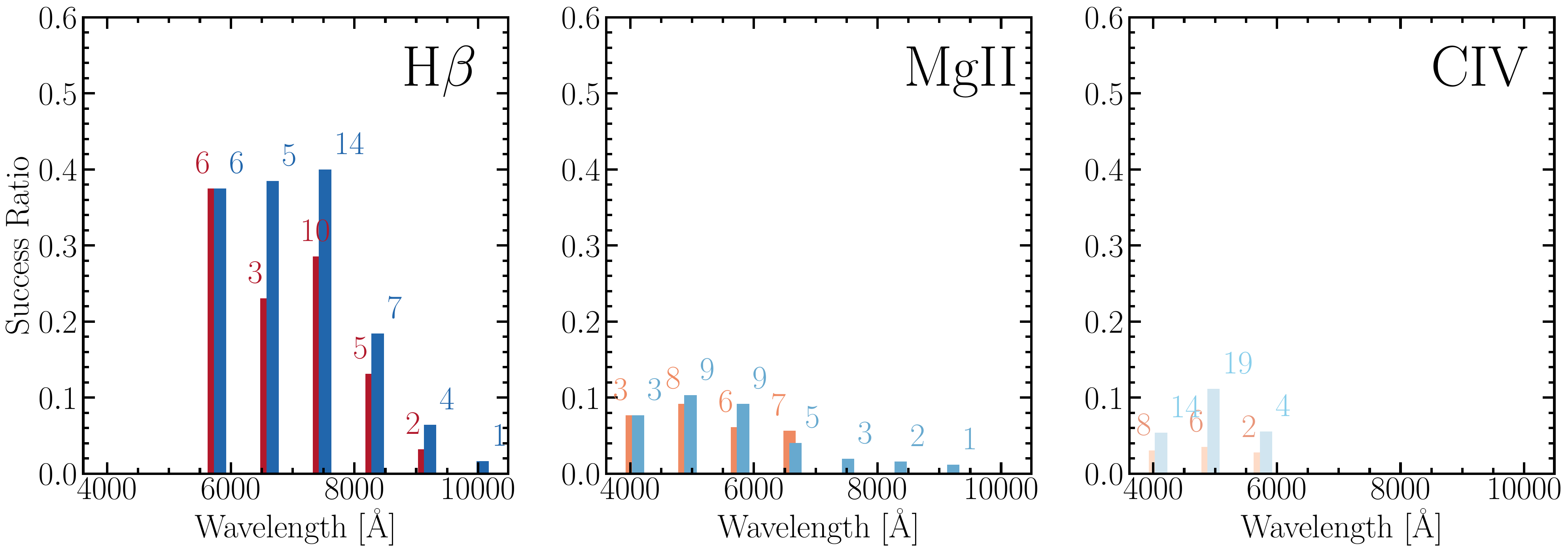}
 \caption{Lag success fraction as a function of emission-line position on the BOSS spectrograph. Early SDSS-RM measurements are depicted in reddish tones, while the 7-year lag measurements are represented by bluish tones. For \Hb\ (left panel), we have the highest ratio of obtaining a successful lag measurement when the emission line is positioned near the center of the detector (middle thirds), and we see a significant decrease in the lag recovery rate as the \Hb\ profile falls on the edges (outer thirds) of the detector both in the single-year \citep{Grier2017} and the 7-year \citep{Shen2024} analyses. The effect of emission-line location on the detector in \ion{Mg}{2} (middle) and \ion{C}{4} measurements (right) are more mixed; this is because the 4-year \ion{Mg}{2} study in \citet{Homayouni2020} has been cut at $z<$1.7 due to contamination with skylines. The \ion{C}{4} emission line can be measured for targets at higher redshift and the lags are plausibly longer than could be recovered with the observed baseline of 4-years \citep{Grier2019} or 7-years \citep{Shen2024}. For each wavelength bin, we also display the number of gold lags to illustrate the statistical significance in each bin.}
\label{fig:detector}
\end{figure*}
The SDSS-RM quasars probe $0.3<z<4.5$; redshift effects limit the observable emission lines that can be targeted with RM observations and cosmological time dilation increases the observed lags.
The throughput of the SDSS BOSS spectrograph exhibits a wavelength dependence across its operational range of 360 nm~$< \lambda <$~1000 nm \citep{Smee2013}. This variation is attributed to several factors, including the CCD sensitivity, the blue/red grism efficiency, the collimator characteristics, and contamination by sky lines. We next explore the potential influence of the targeted emission line's location within the SDSS spectrograph bandpass on the gold-lag measurements. The detector edges exhibit lower sensitivity, resulting in increased noise.  Therefore, we investigate whether the placement of the emission line closer to the detector center improves the probability of acquiring a successful lag measurement.

Considering the redshift range of the SDSS-RM sample (Section~\ref{sec:sdss-rm}), prominent BLR emission lines observable include \Hb\ at lower redshifts, \ion{Mg}{2} in the mid-redshift range, and \ion{C}{4} at higher redshifts. We investigate the impact of target redshift and observed-frame wavelength for each emission line by dividing the bandpass into seven bins of equal observed wavelength range. Figure~\ref{fig:detector} compares the success ratio of the observed-frame emission line on the detector for both early-year SDSS-RM and the 7-year lag measurements. Our analysis reveals a trend of decreasing lag success with wavelength for \Hb, and \ion{Mg}{2}. We also find a significantly higher success rate for the gold \Hb\ lag measurements when the observed frame \Hb\ is located near the mid-wavelength range of the detector. %For \ion{Mg}{2}, the trend is less evident. While the gold \ion{Mg}{2} lag measurements appear slightly more prevalent when the observed-frame emission line falls near the mid-wavelength range, this conclusion is inconclusive for the \ion{Mg}{2} lag measurements by \citet{Homayouni2020} as their study limited the redshift range to $0.35 < z < 1.7$ due to skyline contamination. 
The results for \ion{C}{4} are less conclusive. The highest redshift in the \ion{C}{4} gold sample ($z$ = 2.9) restricts our ability to confirm the lag success rate beyond $\approx$6000 \AA. Additionally, assessing \ion{C}{4} results is complicated as these targets tend to have longer rest-frame lags due to their higher luminosity and cosmological time dilation effects. Our comparison across the three emission lines suggests a clear influence only for \Hb\ where the three central bins covering 5725 \AA\ to 7425 \AA\ exhibit a mean success rate of $\gtrsim$20\%. The central bins from \citet{Shen2024} exhibited a higher success fraction of approximately 39\% compared to the approximately 30\% success rate of the central bins from \citet{Grier2017} and the neighboring bins closer to the spectrograph edges (outer thirds compared to the middle third). The results for \ion{Mg}{2} and \ion{C}{4} were less definitive. 
%Unlike the \Hb\ that is recombination-dominated, the \ion{Mg}{2} line is collisionally excited, which could result to a lower response to continuum variations. The \ion{Mg}{2} lag analysis by \citet{Homayouni2020} revealed that the \ion{Mg}{2} line in contaminated by a lot of skylines in several targets.

%%%%%%%%%%%%%%%%%%%%%%%%%%%%%%%%%%%%%%%%%%%%%%%%%%%%%%%%%%%%%%%%%%%%%%%%%%%%%%%%%%%%%%%%%%
%%%%%%%%%%%%%%%%%%%%%%%%%%%%%%%%%%%%%%%%%%%%%%%%%%%%%%%%%%%%%%%%%%%%%%%%%%%%%%%%%%%%%%%%%%
\section{Light-curve Variability Characteristics} \label{sec:light_curve}
%%%%%%%%%%%%%%%%%%%%%%%%%%%%%%%%%%%%%%%%%%%%%%%%%%%%%%%%%%%%%%%%%%%%%%%%%%%%%%%%%%%%%%%%%%
%%%%%%%%%%%%%%%%%%%%%%%%%%%%%%%%%%%%%%%%%%%%%%%%%%%%%%%%%%%%%%%%%%%%%%%%%%%%%%%%%%%%%%%%%%
\begin{table*}[!t]
\caption{Comparative Analysis of Light Curve Properties: Gold and Parent \label{tab:table3}}
\begin{center}
\begin{tabular}{c|cc|cc|cc|cc}
\hline
\hline
Emission Line & \multicolumn{2}{c}{Frac. RMS Var.}& \multicolumn{2}{c}{SNR2} & \multicolumn{2}{c}{Con. \textit{dw}} & \multicolumn{2}{c}{Line \textit{dw}}\\
{} & \multicolumn{2}{c}{} &  \multicolumn{2}{c}{} &  \multicolumn{2}{c}{} &  \multicolumn{2}{c}{}\\
{} & Parent & Gold & Parent & Gold & Parent & Gold & Parent & Gold \\
\hline
\hline
\Hb\ (first-year) & $0.1^{+0.24}_{-0.06}$ & $0.1^{+0.14}_{-0.05}$ & $10.21^{+10.08}_{-4.79}$ & $13.86^{+15.68}_{-4.1}$ & $1.14^{+0.5}_{-0.58}$ & $0.55^{+0.49}_{-0.3}$ & $1.42^{+0.42}_{-0.58}$ & $0.66^{+0.4}_{-0.19}$ \\ [8pt]
MgII (four-year) & $0.1^{+0.06}_{-0.04}$ & $0.1^{+0.04}_{-0.03}$ & $29.48^{+11.67}_{-6.7}$ & $25.88^{+11.71}_{-3.48}$ & $0.56^{+0.61}_{-0.33}$ & $0.44^{+0.58}_{-0.22}$ & $1.16^{+0.35}_{-0.44}$ & $1.08^{+0.36}_{-0.55}$ \\ [8pt]
CIV (four-year) & $0.11^{+0.07}_{-0.05}$ & $0.14^{+0.09}_{-0.06}$ & $32.96^{+18.49}_{-9.46}$ & $42.34^{+15.18}_{-10.86}$ & $0.69^{+0.53}_{-0.38}$ & $0.46^{+0.23}_{-0.26}$ & $0.91^{+0.42}_{-0.39}$ & $0.49^{+0.29}_{-0.18}$ \\ [8pt]
\Hb\ (seven-year) & $0.16^{+0.25}_{-0.08}$ & $0.18^{+0.14}_{-0.09}$ & $23.98^{+28.77}_{-9.69}$ & $51.4^{+27.15}_{-19.9}$ & $0.62^{+0.43}_{-0.35}$ & $0.47^{+0.39}_{-0.25}$ & $1.0^{+0.52}_{-0.57}$ & $0.39^{+0.38}_{-0.19}$ \\ [8pt]
MgII (seven-year) & $0.1^{+0.07}_{-0.04}$ & $0.1^{+0.13}_{-0.04}$ & $22.33^{+14.88}_{-8.6}$ & $36.3^{+14.62}_{-11.57}$ & $0.67^{+0.41}_{-0.36}$ & $0.38^{+0.38}_{-0.2}$ & $1.39^{+0.31}_{-0.42}$ & $0.78^{+0.38}_{-0.37}$ \\ [8pt]
CIV (seven-year) & $0.12^{+0.08}_{-0.05}$ & $0.13^{+0.06}_{-0.07}$ & $33.09^{+21.39}_{-14.07}$ & $46.7^{+28.63}_{-13.8}$ & $0.67^{+0.42}_{-0.33}$ & $0.6^{+0.37}_{-0.27}$ & $1.05^{+0.38}_{-0.41}$ & $0.64^{+0.37}_{-0.21}$ \\ [8pt]
\hline
\hline
\end{tabular}
\end{center}
\end{table*}
\begin{figure*}
\centering
\includegraphics[width=0.85\textwidth]{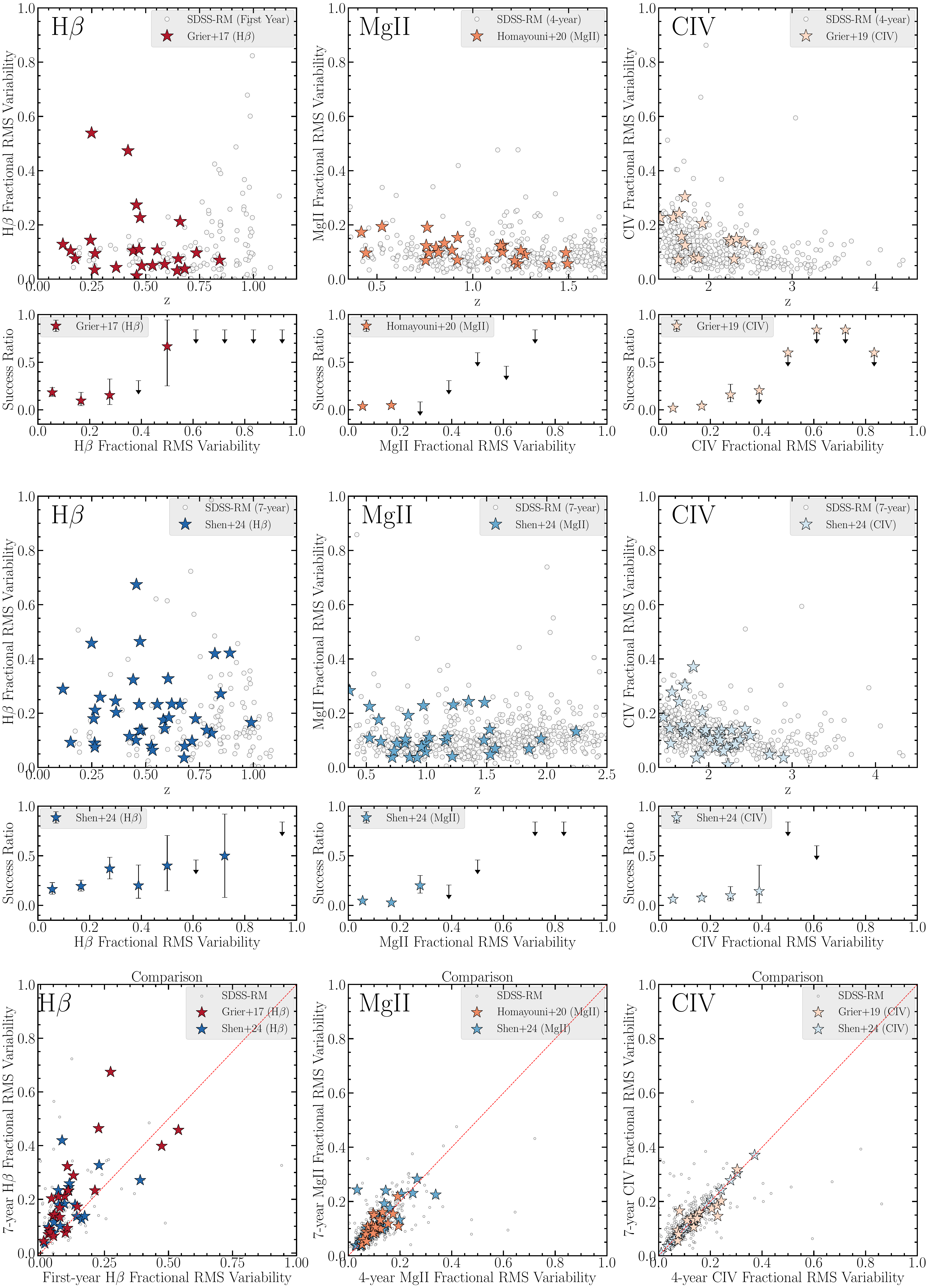}
 \caption{Fractional root-mean-square (rms) variability for \Hb\,, \ion{Mg}{2}, and \ion{C}{4} as a function of redshift for early and 7-year SDSS-RM results (see Figure~\ref{fig:ew} for the description of layouts, symbols, and legends). The gold-lag success fraction as a function of fractional RMS variability  reveals no discernible trends for the corresponding emission line (see Table~\ref{tab:table3} for a comparison of median values in each work). The bottom panels compare the fractional rms variability between the early-year and 7-year SDSS-RM campaigns. Our analysis reveals no statistically significant correlation between the emission line fractional rms variability and gold-lag measurements.}
\label{fig:frac_var}
\end{figure*}
%\Hb\ (first-year) & $0.10^{+0.33}_{-0.04}$ &  $0.1^{+0.23}_{-0.04}$ & $10.21^{+20.29}_{-5.43}$ & $13.86^{+29.54}_{-9.76}$ & 0.6 $\pm$ 0.05 & 0.3 $\pm$ 0.1 & 1.4 $\pm$ 0.1& 0.8 $\pm$ 0.1 \\ 
%MgII (four-year) & $0.10^{+0.15}_{-0.06}$ & $0.1^{+0.14}_{-0.07}$ & $29.48^{+41.15}_{-22.78}$ & 25.9 $\pm$ 2.0 & 0.4 $\pm$ 0.04 & 0.4 $\pm$ 0.1 & 1.2 $\pm$ 0.04 & 1.0 $\pm$ 0.1\\
%CIV (four-year) & $0.11^{+0.18}_{-0.06}$ & $0.14^{+0.23}_{-0.08}$ & $32.96^{+51}_{-0.04}$ & 42.3 $\pm$ 4.7 & 0.5 $\pm$ 0.04 & 0.4 $\pm$ 0.1 &  1.0 $\pm$ 0.02 & 0.5 $\pm$ 0.1\\
%\Hb\ (seven-year) & $0.16^{+0.40}_{-0.08}$ & $0.18^{+0.32}_{-0.09}$ & 24.0 $\pm$ 1.6 & 51.4 $\pm$ 5.3 & 0.6 $\pm$ 0.04 & 0.5 $\pm$ 0.1 & 1.0 $\pm$ 0.1 & 0.4 $\pm$ 0.1\\
%MgII (seven-year) & $0.10^{+0.17}_{-0.06}$ & $0.10^{+0.23}_{-0.06}$ & 22.3 $\pm$ 0.5 & 36.3 $\pm$ 3.0 & 0.7 $\pm$ 0.02  & 0.4 $\pm$ 0.1 &  1.4 $\pm$ 0.01 & 0.8 $\pm$ 0.1\\
%CIV (seven-year) & $0.12^{+0.19}_{-0.07}$ & $0.13^{+0.19}_{-0.06}$ & 33.1 $\pm$ 0.8 & 46.7 $\pm$ 4.3 & 0.7 $\pm$ 0.03 & 0.6 $\pm$ 0.1 &  1.0 $\pm$ 0.02 & 0.6 $\pm$ 0.1\\

\subsection{Fractional Variability}
Quasars exhibit variability characterized by a wide range of amplitudes, wavelengths, and timescales (e.g., \citealp{Collier2001, Peterson2004, Kelly2009, MacLeod2012}). This variability is crucial for RM studies. However, not all quasars exhibit sufficient variability of their emission lines to enable reliable RM measurements. It is plausible that the success of lag measurements is linked to specific characteristics of the light-curve variability. \citet{Shen2024} quantified the intrinsic variability of SDSS-RM light curves using a maximum-likelihood estimator. They report the intrinsic root-mean-square (rms) variability for both the 11-year photometric light curve and the 7-year emission-line light curves. Their findings indicate that different emission lines exhibit varying degrees of intrinsic variability, although a general correlation exists between the continuum and emission-line rms variability.

We investigate the connection between emission-line rms variability and the success rate of gold-lag measurements. We utilize the fractional rms variability (normalized to the mean flux) measured by PrepSpec as described in \citet{Shen2019a}. Our analysis reveals a broad distribution of fractional rms variability within the gold sample. %%Several objects have fractional rms variability exceeding the sample median, \q{where the median fractional RMS variability in the parent and the gold sample is $\approx 0.1\pm 0.01$ in the early-year and the 7-year SDSS-RM lags}, although the mean fractional RMS variability in both the parent and the gold sample of \Hb\ lags of \citet{Shen2024} are higher $0.18 \pm 0.03$ and $0.16 \pm 0.01$. This discrepancy could be attributed to the presence of multiple random fluctuations within the light curve, which affects the identification of a reliable lag. This is particularly evident for the \Hb\ emission line, where 
Significant sky line residuals contaminate the \Hb\  fractional RMS variability, potentially leading to overestimated values \citep{Shen2019a}. While the effect of skylines is less pronounced for \ion{Mg}{2}, it is still present at $z>1.5$ among the lag measurements of \citet{Shen2024}, where skylines contaminate the spectra. The \ion{Mg}{2} lag measurements in \citet{Homayouni2020} were cut at a redshift of $z$ = 1.7 due to the contamination by skylines, therefore, there are no lag measurements for \ion{Mg}{2} at $z > 1.7$ based on the early SDSS-RM study \citep{Homayouni2020} and the 7-year measurements reported by \citet{Shen2024} only includes three targets at z$>1.7$. 

The results of this investigation are presented in Figure~\ref{fig:frac_var}. We further compared the fractional rms variability between the early-year and 7-year studies for each emission line in the bottom panels of Figure~\ref{fig:frac_var}. It is noteworthy that the 4-year and 7-year results largely follow a 1:1 trend. The deviation observed in the comparison between the 1-year and 7-year \Hb\ fractional rms variability (bottom left corner of Figure~\ref{fig:frac_var}) can be attributed to the variability of quasars on the observed-frame timescales. The damping timescale of quasars are longer than the seasonal monitoring duration.

\begin{figure*}
\centering
\includegraphics[width=0.9\textwidth]{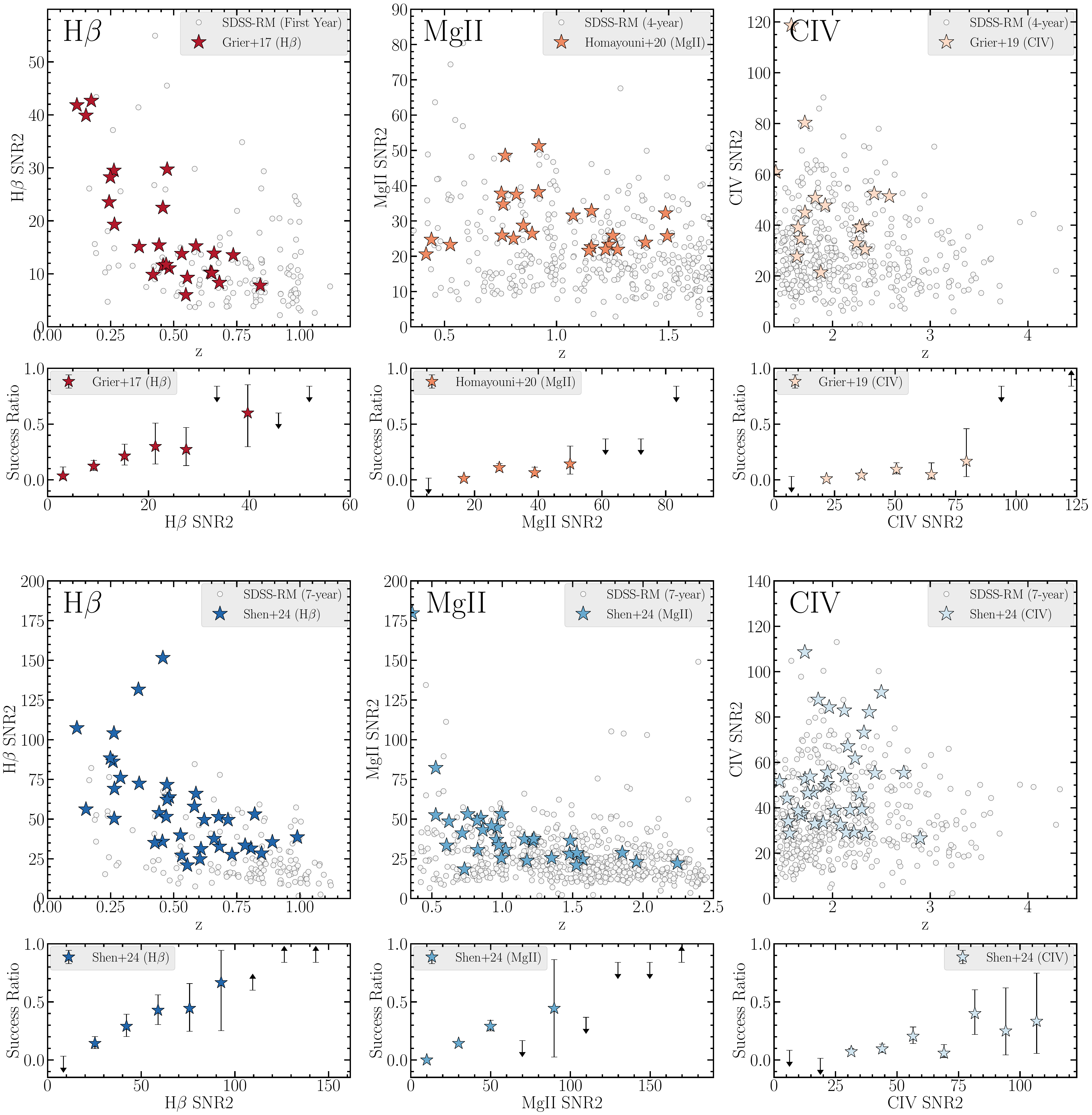}
 \caption{Emission-line variability signal-to-noise ratio (SNR2) as a function of redshift for the SDSS-RM survey (see Figure~\ref{fig:ew} for the description of layouts, symbols, and legends). Only the lag measurements by \citet{Grier2019} and \citet{Homayouni2020} adopted a SNR2 threshold for the initial target selection in lag measurements. Our analysis reveals that SNR2 has a positive trend with gold lag success fraction. This is more evident in the \Hb\ and \ion{C}{4} lines, especially in the 7-year analysis compared to the earlier lag measurements. Overall, SNR2 generally exhibits a more stable indicator of gold-lag measurements when observed for multi-season campaigns. Nevertheless we find significant scatter for the SNR2 and the underlying parent sample. Table~\ref{tab:table3} provides a comparison of median SNR2 values in each work.}
\label{fig:snr2}
\end{figure*}

\subsection{SNR2}
Light-curve variability can also be quantified using an empirical variability metric known as the variability signal-to-noise ratio (SNR2). This metric, as computed by PrepSpec, is calculated as the square root of the chi-squared statistic ($\chi^2$) minus the degrees of freedom (DOF), expressed as $N$-1, where $N$ represents the number of data points in the light curve ($\rm SNR2 = \sqrt{\chi^2 - DOF}$) and the $\chi^2$ is calculated against the mean flux. Therefore, smaller $\rm SNR2$ indicates that the light curve is not variable compared to the mean flux and higher $\rm SNR2$ values indicate a light curve with greater intrinsic variability where the null hypothesis signifies a poor fit for a constant light curve.

To enhance the efficiency of RM lag analysis, some studies have proposed pre-selecting targets based on the variability of their emission-line light curves. This approach aims to reduce the number of unlikely candidates for lag detection. Early SDSS-RM studies, such as those by \citet{Grier2019} and \citet{Homayouni2020}, employed a threshold of SNR2 $>$ 20 for their initial, larger parent samples. However, this practice was not universally adopted among the SDSS-RM work, potentially introducing bias into some early SDSS-RM results.

%Some studies have advocated for pre-selecting targets based on emission line light curve variability before implementing a detailed RM lag analysis to reduce the number of unlikely candidates for lag detection. Some early-year SDSS-RM studies, such as those by \citet{Grier2019} and \citet{Homayouni2020}, employed a threshold of SNR2$ >$ 20 on their initial, larger parent samples. However, this practice was not adopted in all SDSS-RM lag measurements, potentially introducing bias into some early-year SDSS-RM results. Despite this inconsistency, we investigate the potential connection between SNR2 and the success in gold lag measurements. The 7-year study by \citet{Shen2024} found a direct connection between the SNR2 values and lag detection rate (not necessarily restricted to the gold sample, but for the overall lag measurements) for the \Hb\ sample compared to \ion{Mg}{2} and \ion{C}{4}, where the detection rate increased with SNR2 but at a slower rate (see Figure 16 of \citealt{Shen2024}). 

Both \ion{C}{4} and \ion{Mg}{2} in \citet{Grier2019} and \citet{Homayouni2020} studies employed an SNR2 $>$ 20 threshold in their parent samples, whereas the lag measurements reported in \citet{Grier2017, Shen2024} did not. Notably, a strong preference for this threshold exists even though it was not consistently applied across all studies (Figure~\ref{fig:snr2}). While \citet{Shen2024} found a 70\% lag detection rate for \Hb\ for targets with SNR2 $>$ 35, a similar analysis of SNR2 in \citet{Grier2017} revealed that the majority of gold lags had SNR2 $<$ 20; however, this might also be connected to the higher cadence during the first-year of SDSS-RM observation (see Section~\ref{sec:cadence} for a thorough discussion).

To investigate the potential relationship between SNR2 and the success of gold-lag measurements, we examined the lag measurements by \citet{Grier2017, Grier2019}, \citet{Homayouni2020}, and \citet{Shen2024}. Our investigation revealed a direct correlation between SNR2 values and the gold-lag detection rate for the \Hb\ sample compared to \ion{Mg}{2} and \ion{C}{4}. While the detection rate somewhat increases with higher SNR2 for all three emission lines, the increase was more pronounced for \Hb. \citet{Shen2024} observed a comparable trend for all lag measurements in their dataset, regardless of whether they were classified as gold (see Figure 16 of \citet{Shen2024}). Similar to fractional RMS variability, SNR2 inherently increases with longer light-curve baselines. We therefore refrain from showing comparisons between one-year, 4-year, and 7-year analyses due to the duration and the parent sample selection criteria in Figure~\ref{fig:snr2}. These results indicate, although SNR2 can be a valuable preliminary screening metric, its effectiveness is influenced by the cadence and duration of monitoring.

\subsection{Durbin-Watson Statistic}
While our findings suggest the importance of variability as measured by SNR2, instances exist where a gold lag remains undetected despite relatively high SNR2 values. For time-series analysis software to effectively detect an RM lag, the light curve often requires a distinct ``hook" or inflection feature in the continuum and line light curves. In this section, we investigate whether the presence of serial correlations in the continuum and emission line light curve would lead to a higher success in detecting a gold-lag measurement. Serial correlation refers to the dependence of serial data points, indicating a patterned behavior in the time series. A commonly used test for first-order autocorrelation, which assumes independent error terms, is the Durbin-Watson test (Durbin \& Watson 1950). The Durbin–Watson test statistic is given by

%To do this, we consider how well the residuals from a regression model can be predicted by the residuals from the previous period. This would explore if it there is a ``pattern" present in the time-series. A popular test for this first order autocorrelation is the Durbin-Watson test (DW, \yh{Durbin and Watson 1950}), which assumes that the all error terms are independent.
\begin{equation}
    dw = \frac{\Sigma_{t = 2}^{T} (e_t - e_{t-1})^2}{\Sigma_{t = 1}^{T} e_t^2} \approx 2 - 2r,
\end{equation}
%which translate to $dw \approx 2 - 2r$, where $r$ is the first order autocorrelation coefficient.
where $e_t$ is the least-square residual and $r$ is the first order autocorrelation coefficient. Consequently, a value of $ dw\approx2$ indicates that the first order autocorrelation coefficient $r\approx 0$. If $dw < 2$, this is an indication for positive autocorrelation $r > 0$; if $dw > 2$, then $r<0$. Therefore, we can assess the connection between the autocorrelation, as indicated by the $dw$ statistics, which can identify ``hooks" in the light curves and might serve as a reliable predictor of the gold-lag measurements.

\begin{figure*}
\centering
\includegraphics[width=0.8\textwidth]{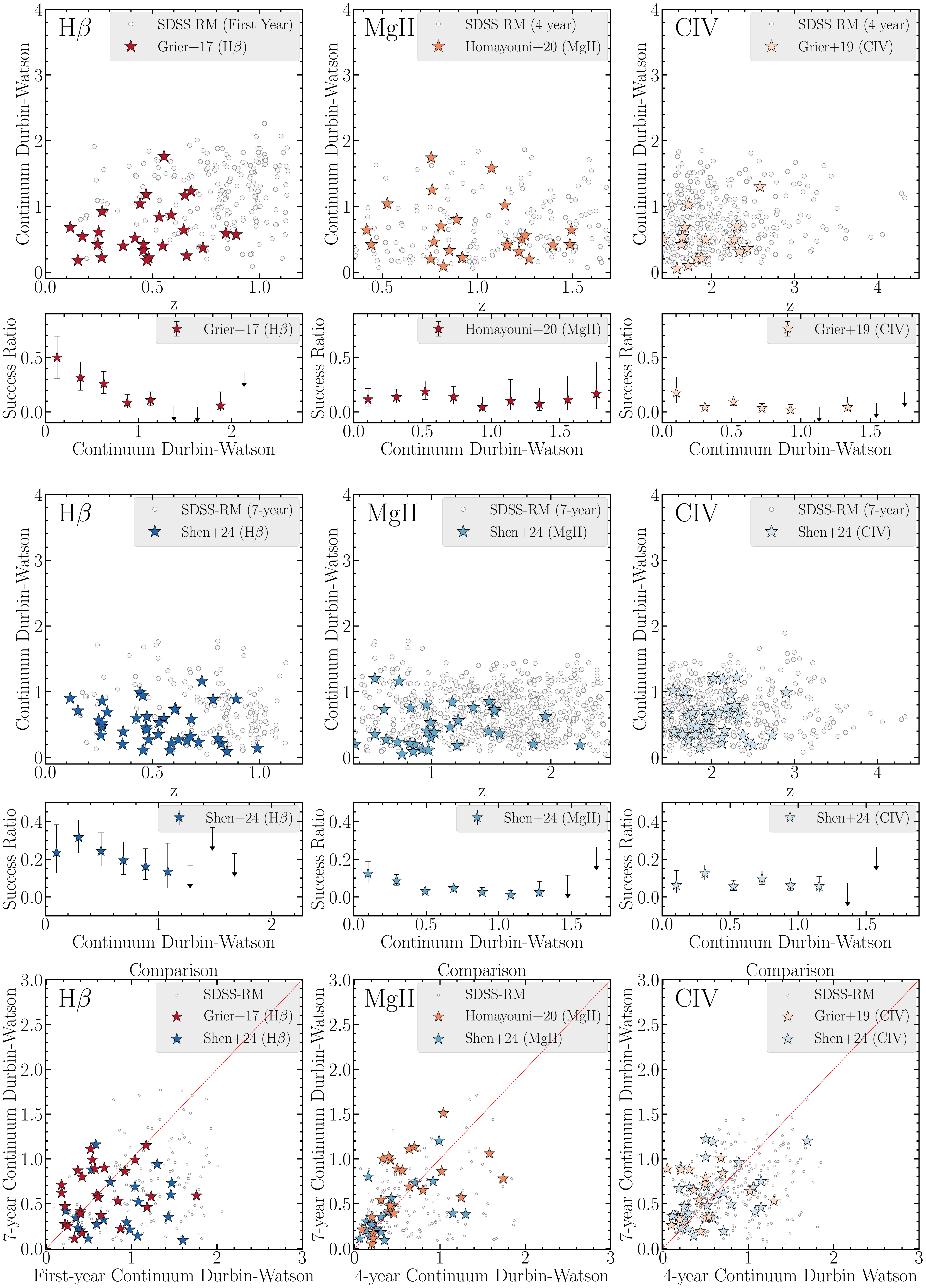}
 \caption{The Durbin-Watson statistic computed for the continuum light curves in \Hb, \ion{Mg}{2}, and \ion{C}{4} (see Figure~\ref{fig:ew} for the description of layouts, symbols, and legends). The Durbin-Watson statistic shows the auto-correlation of the continuum light curve with a lag of one epoch. Positive serial auto-correlation ($r$ = 1) has a $dw$ = 0, no serial auto-correlation has a $dw$~=~2, and negative serial auto-correlation has a $dw$ = 4. The bottom row, compares the 7-year vs. the early-year SDSS-RM continuum Durbin-Watson statistics. We see an abundance of successful lag measurements for continuum light curves that have $dw<1$. While generally the gold-lag success rate does not exhibit a strong correlation with continuum light curve \textit{dw}, a trend is more robustly present for the \Hb\ lags in early-year SDSS-RM results \citep{Grier2017}. This aligns with logistic regression analyses where the continuum Durbin-Watson statistic was identified as a robust indicator for gold-standard lag measurements (\textit{p} $<$ 0.001; \citealt{Grier2017}) and, albeit less strongly, for \ion{C}{4} lags (\textit{p} = 0.045; \citealt{Grier2019}), proving generally more predictive for these lags than for the other studies. A comparison of median \textit{dw} values in Table~\ref{tab:table3} reveals that the continuum \textit{dw} of gold-lag targets is generally comparable to that of the parent population.}
\label{fig:dw_con}
\end{figure*}

\begin{figure*}
\centering
\includegraphics[width=0.86 \textwidth]{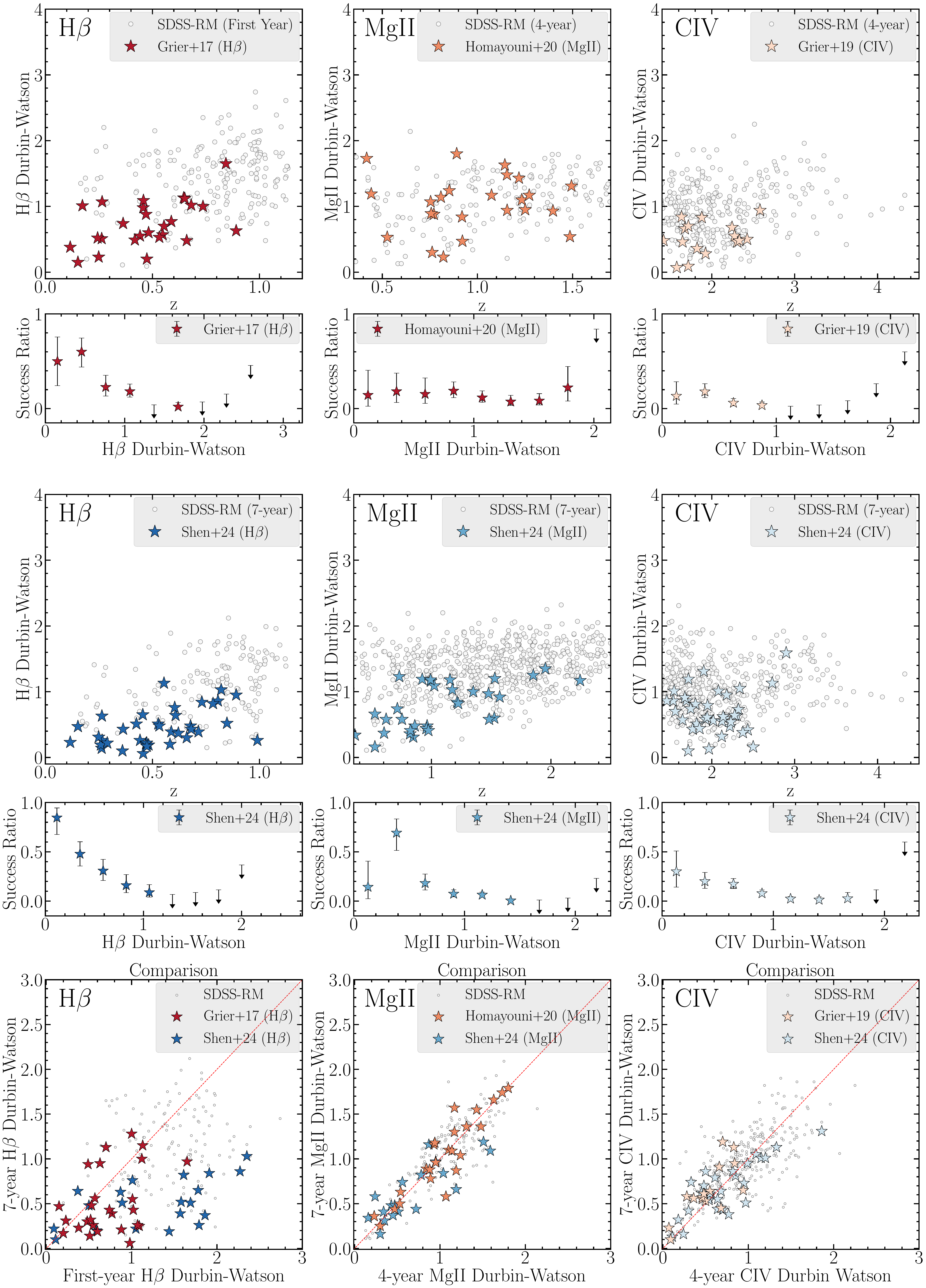}
 \caption{Similar to Figure~\ref{fig:dw_con} but for the \Hb\ (left column), \ion{Mg}{2} (middle column), and \ion{C}{4} (right column) light curves. Similar to Figure \ref{fig:dw_con}, we see that the majority of successful lag measurements correspond to light curves with  $dw<$1. The gold-lag success ratio is significantly correlated with emission-line $dw$ for values less than 1. Comparative analyses indicate that the emission line $dw$ of gold-lag targets is consistently lower than for the overall population (see Table~\ref{tab:table3}), with a more pronounced effect for \Hb\ and a less significant one for \ion{Mg}{2}.}
\label{fig:dw_line}
\end{figure*}

We illustrate the result of this comparison in Figures~\ref{fig:dw_con} and \ref{fig:dw_line} for the continuum and emission-line light curve respectively. The \textit{dw} statistic, particularly for the emission-line light curves, exhibits a strong preference for values below 1 within the gold lag sample (See Figure~\ref{fig:dw_line}). The trend toward $dw\,<$ 1 in the gold lag sample is particularly evident in the 4-year \ion{C}{4} lag measurements of \citet{Grier2019} and the 7-year results reported by \citet{Shen2024}. This suggests that a $dw < 1$ might be an indicator for successful gold lag detection. The \Hb\ lags of \citet{Grier2017} exhibit a less pronounced preference for $dw <1$, potentially indicating that this threshold might be more applicable for longer-term observations.  Furthermore, the gold \ion{Mg}{2} lag measurements in \citet{Homayouni2020} display a wider \textit{dw} distribution, which could be related to the weaker response of the \ion{Mg}{2} emission line to continuum variations. The continuum \textit{dw} values generally exhibit a broader distribution, with most still falling below 1.5 (see Figure~\ref{fig:dw_con}). This indicates that the $dw$ derived from the emission-line light curve is a better indicator of gold-lag measurements  than $dw$ computed from the continuum light curve. 

Figures~\ref{fig:dw_con} and \ref{fig:dw_line} show the distribution of the continuum light curve \textit{dw} and emission line \textit{dw} in comparison to the SDSS-RM \textit{dw} distribution. %To further investigate this link, we compare the \textit{dw} in both the continuum and emission line light curves simultaneously to targets where both the continuum and emission line dw values are less than 1. \q{This analysis recovers 61\%, 41\%, 81\% of gold lags in \Hb\ \citep{Grier2017}, \ion{Mg}{2} \citep{Homayouni2020}, and \ion{C}{4} \citep{Grier2019}}. The recovery ratio of gold lags is higher based on the 7-year gold lags of \citet{Shen2024}  89\%, 62\%, 67\%  in \Hb, \ion{Mg}{2}, and \ion{C}{4} respectively, which could indicate that the Durbin-Watson statistics is a better indicator when targets are observed on longer baseline.

Thus far, we have focused on the intrinsic properties of the quasars and their light curves, investigating how these characteristics influence the success rate of RM lag measurements. While target selection and variability are crucial aspects, another important factor in RM studies is the cadence of the observations. Given the lag measurements from the SDSS-RM observation, we now focus on the observational cadence and what are the implications for future RM campaigns.

%\begin{figure*}
%\centering
%\includegraphics[width=0.8\textwidth]{fig5_con_dw_hist.pdf}
% \caption{Another representation of the Durbin-Watson Statistics computed for the continuum light curve in early-year SDSS-RM (left panels) and the 7-year SDSS-RM results (middle panels).}
%\label{fig:dw_con_hist}
%\end{figure*}

%The middle panels show two example of continuum light curve with D$-$W$<$1 where the light curve shows sequential correlated variability. The right panels show two example light curves with D$-$W $>$ 1, where the light curve does not show any clear features suitable for lag analysis. In each panel, the red dashed line shows the median of the light curve.

%\begin{figure*}
%\centering
%\includegraphics[width=\textwidth]{fig6_line_dw_hist.pdf}
% \caption{Similar to Figure~\ref{fig:dw_line} but showing the distribution for the \Hb\ (top row), \ion{Mg}{2} (middle row), and \ion{C}{4} (bottom row) light curves. Similar to \ref{fig:dw_con}, we see that the majority of successful lag measurements correspond to \Hb\ light curves with dw $<$1.}
%\label{fig:dw_line_hist}
%\end{figure*}

%%%%%%%%%%%%%%%%%%%%%%%%%%%%%%%%%%%%%%%%%%%%%%
%%%%%%%%%%%%%%%%%%%%%%%%%%%%%%%%%%%%%%%%%%%%%%
\subsection{Logistic Regression}\label{sec:log_reg}
%%%%%%%%%%%%%%%%%%%%%%%%%%%%%%%%%%%%%%%%%%%%%%
%%%%%%%%%%%%%%%%%%%%%%%%%%%%%%%%%%%%%%%%%%%%%%
To further investigate which of the target properties most significantly influence the distribution of gold-lag measurements, we conducted a logistic regression for each of the six distinct sets of lag measurements. The independent variables considered in our model were \lum, REW, fractional variability, SNR2 variability, and the continuum and line Durbin-Watson statistics. Our analysis of the early SDSS-RM lag measurements reveals that for the \Hb\ lag measurements reported by \citet{Grier2017}, the \lum, along with both the continuum and line Durbin-Watson statistics, were the strongest predictors of a gold-lag measurement. However, a similar analysis for the \ion{Mg}{2} lags in \citet{Homayouni2020} revealed only a marginal predictive power for \lum ($p$=0.048). For the CIV lags reported in \citet{Grier2019}, the line Durbin-Watson statistic ($p$=0.003) and, with less significance, the continuum Durbin-Watson statistic ($p$=0.045) contributed to the identification of gold lags. For the 7-year lag measurements reported by \citet{Shen2024}, our analysis indicates that for the \Hb, as well as the \ion{C}{4} results, \lum\ and the line Durbin-Watson statistic were identified as the strongest predictors. On the other hand, for the \ion{Mg}{2} lags, the line Durbin-Watson statistic was identified as the single robust predictor of gold-lag measurements. 

We also note that while the logistic regression reveals a negative correlation (as also illustrated in Figure~\ref{fig:luminosity}), this trend may stem from inherent biases within our sample and the observational constraints of these studies. Specifically, the inverse connection to \lum\ may reflect longer time delays in brighter sources (e.g., \citealt{Bentz2013}) that demand longer observational baselines not fully covered by these studies. With the exception of the \ion{Mg}{2} line, the line Durbin-Watson statistic consistently emerged as the strongest predictor for identifying gold lags across all other emission lines examined in this work.

\vspace{5mm}

%%%%%%%%%%%%%%%%%%%%%%%%%%%%%%%%%%%%%%%%%%%%%
%%%%%%%%%%%%%%%%%%%%%%%%%%%%%%%%%%%%%%%%%%%%%
\section{Cadence Properties} \label{sec:cadence}
%%%%%%%%%%%%%%%%%%%%%%%%%%%%%%%%%%%%%%%%%%%%%
%%%%%%%%%%%%%%%%%%%%%%%%%%%%%%%%%%%%%%%%%%%%%

\subsection{Cadence Intensity in SDSS-RM}

A main factor in RM observations is the cadence, as each epoch is valuable. Given the limited number of epochs available due to telescope time and weather constraints, RM programs must optimize their cadence to maximize lag measurement success. Simulations by \citet{Shen2015a} demonstrated the advantages of high-cadence observations for lag detection. While their work provided valuable guidance for the SDSS-RM initial cadence design, it had two limitations: 1) it was based on a single observation season, and 2) it used simulated light curves, which may not fully represent real-observing conditions and lag measurement success rates. To address these shortcomings, the current work examines the effect of reduced cadence using actual SDSS-RM data. This is done by comparing the impact of a relatively uniform cadence to the denser SDSS-RM cadence in the initial year (2014) and a sparser cadence in subsequent years (2015 - 2020). The SDSS-RM project, spanning seven years, employed a varying cadence design. In 2014, spectroscopy occurred every four days on average, totaling 32 epochs. From 2015 to 2017, two epochs were captured monthly, averaging 12 epochs annually. During 2018-2020, the cadence decreased to six epochs per year. Overall, the project produced a spectroscopic light curve with 90 epochs over seven years.

\begin{figure*}
\includegraphics[width=1.0\linewidth]{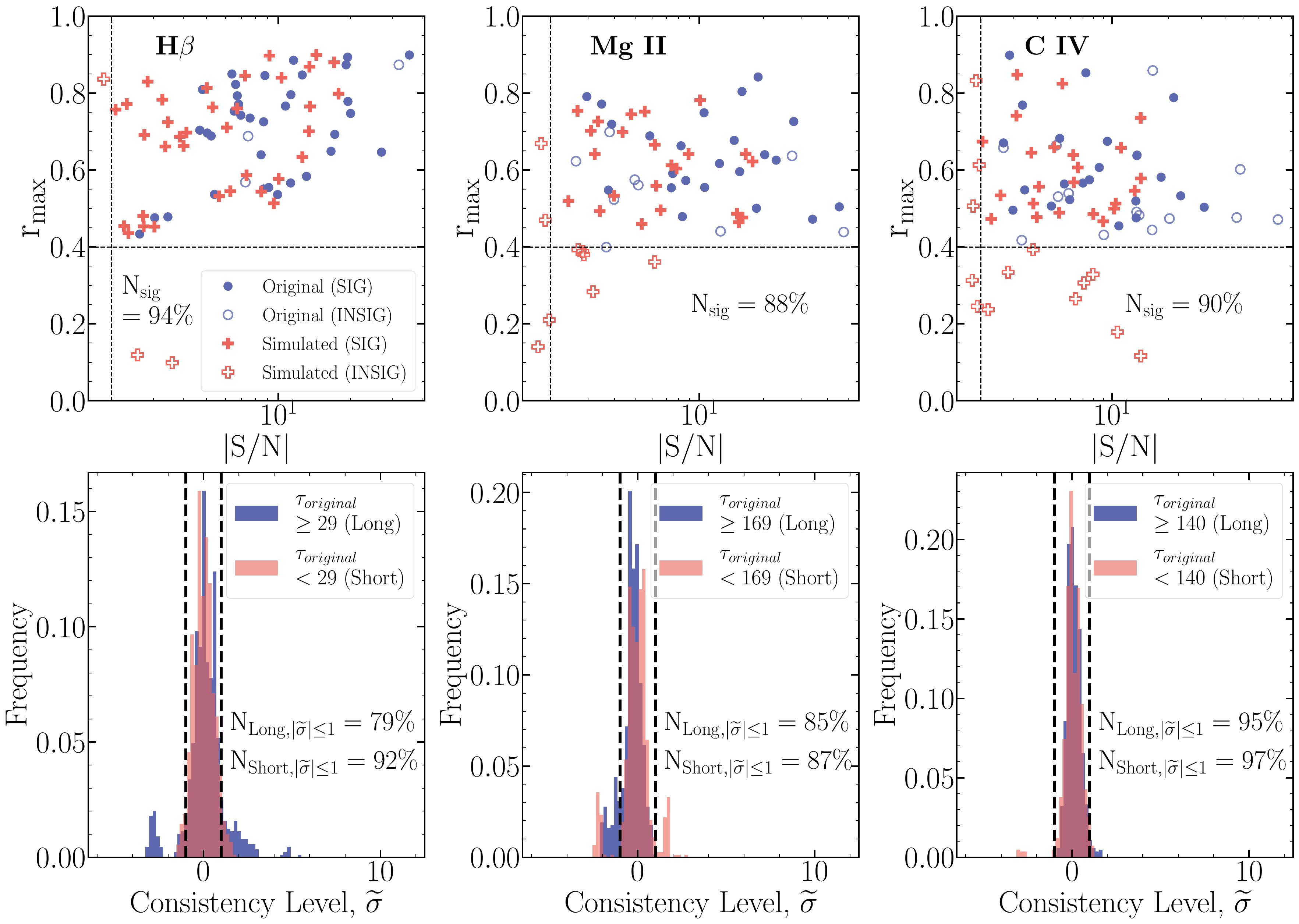}
 \caption{Summary of the simulated lag measurements from the sample of H$\beta$ (left), \ion{Mg}{2} (middle) and \ion{C}{4} (right) from the gold results in \citet{Shen2024}. For each emission line, the top panels shows the maximum change in the statistical criteria in the simulated light curves. Here the dotted points represent the original statistical values on $r_{\rm max}$ and lag difference from zero, while the crosses represent the minimum statistical values among the 50 simulations for each target. The targets for which all cadence-reduced simulations are still significant are marked in blue, while the targets for which at least one cadence-reduced simulation is insignificant are marked in red. In each panel, $N_{\rm sig}$ reports the ratio of statistically significant simulations to the number of all simulations. Overall, 94\% of the \Hb, 88\% of \ion{Mg}{2}, and 90\% of the \ion{C}{4} simulations are still statistically significant. The bottom panel illustrates the distribution of the consistency level between the original and simulated lag for long and short-lag groups (separated by the median original lag of each line). Here the dashed line identify the 1$\sigma$ consistency level. The ratio of consistent lags (to within 1$\sigma$) are reported in each panel. Overall, if we require the simulated lags are still statistically significant and consistent with original lags in $\pm 1\sigma$, we can still recover 81\%, 76\%, and 86\% of all simulations for \Hb, \ion{Mg}{2}, and  for \ion{C}{4} respectively.} %The right panels show the distribution of relative difference $\delta$ between the original and simulated lags divided for the short and long lag measurements. The dashed lines identify the 34\% threshold of the relative difference. The ratios of lags that have a relative difference of $|\delta|\le 0.34$ are reported for the short and long-lag samples. }  %Overall, our simulations shows that by reducing the cadence in the first season, we can still recover more than 85\% of the statistically significant lag measurements, even within the short-lag sample, to within $1\sigma$.% 
 %The $N_{Long,|\sigma|\leq1}\%, N_{Short,|\sigma|\leq1}\%, N_{Sig,|\sigma|\leq1}\%$ are the ratios about simulations within $1 \sigma$ errors of all significant results in the long-lag group, short-lag group and all simulations, respectively.
\label{fig:sim_lag}
\end{figure*}

%Our first experiment is to reduce the cadence and remove epochs from the real data to see if we can still recover a significant and similar lag. We select targets in the gold sample of the SDSS-RM to perform the investigation. We reduce the density of epochs for line light curves in the first year (2014) to 40\%, keeping the epochs and observations the same in the subsequent years. %By doing this, we can get a similar number of epochs in the following 3 years (2015-2017) for line light curves and remain at about 78\% of the total number of epochs. 
To assess the impact of reduced cadence, our  experiment is to select the light curves from the gold sample of \citet{Shen2024} and decrease the epoch density for line light curves in the initial year by 40\% so that the first year has $\sim$13 epochs. We maintained the same number of epochs as the actual observed in subsequent years ($\sim$13 epochs). To implement this, we first analyzed the interval distribution between epochs in 2015-2017, finding a mean of 12 days. Employing this distribution as a prior, we removed epochs at random until a subset of 13 remained. Epochs with shorter intervals, less than 12 days, were preferentially eliminated compared to those with longer cadences. This process simulated the impact of weather loss or other unforeseen events. We repeated this procedure 50 times for each target, generating 50 simulated light curves per quasar in the gold sample. This closely resembled real observations while maintaining a reduced cadence similar to that in the later years.

%To produce the simulated light curves as similar to the real observations as possible, we start with the gold sample of \citet{Shen2024}. We investigate the distribution of intervals between one epoch and the next in 2015-2017 because these three years have the same cadence and can give a statistically significant distribution. The final distribution of intervals is centered around 12 days in 2015-2017. We then take this distribution as a prior to remove the epochs randomly from the first year of data. The average cadence of 12 days in 2015-2017 indicates that we are more likely to keep two epochs in the first season with a 12-day interval after removing the epochs randomly based on the prior distribution of intervals in 2015-2017. However, other intervals can still be chosen but with less probability. This strategy can mimic the true observations well, considering the weather loss or other accidental events. We then repeat this random process 50 times on line light curves to obtain 50 simulated light curves for each target in the gold sample. 

Finally, we run \texttt{PyROA} for the simulated line light curves with their original continuum light curves and apply lag identification and alias removal process to get the final lag measurement. The reason we keep the original continuum light curves and do not remove epochs from them is that the photometric observations for continuum light curves are not as expensive to make as the spectroscopic observations and are made by telescopes worldwide, so the continuum light curves are therefore densely sampled. Specifically, we have 800 epochs along 11 years and 158 epochs in 2014 (the first year of line light curves) on average, which is extremely dense with a cadence of 3 days per epoch approximately. We also perform some tests on running \texttt{PyROA} for cadence reduced light curves for both continuum and emission lines. The distribution of interval between one epoch to the next is almost uniform for each year, i.e., we also reduced 20\% of the total epochs for continuum light curves of several targets selected randomly from the gold sample and found similar lag PDFs using \texttt{PYROA}\textbf{.} %\yz{Comments: I don't recall exactly, but I believe it doesn't imply a uniform continuum light curve. Rather, epochs were randomly removed based on a uniform distribution, with 20\% of epochs removed in each realization.} 
Therefore, for convenience, removing epochs in the first year in the following text refers to randomly reducing the density of epochs in the first year of the original emission line light curve to 40\% unless otherwise specified.

\begin{figure}
\includegraphics[width=0.5\textwidth]{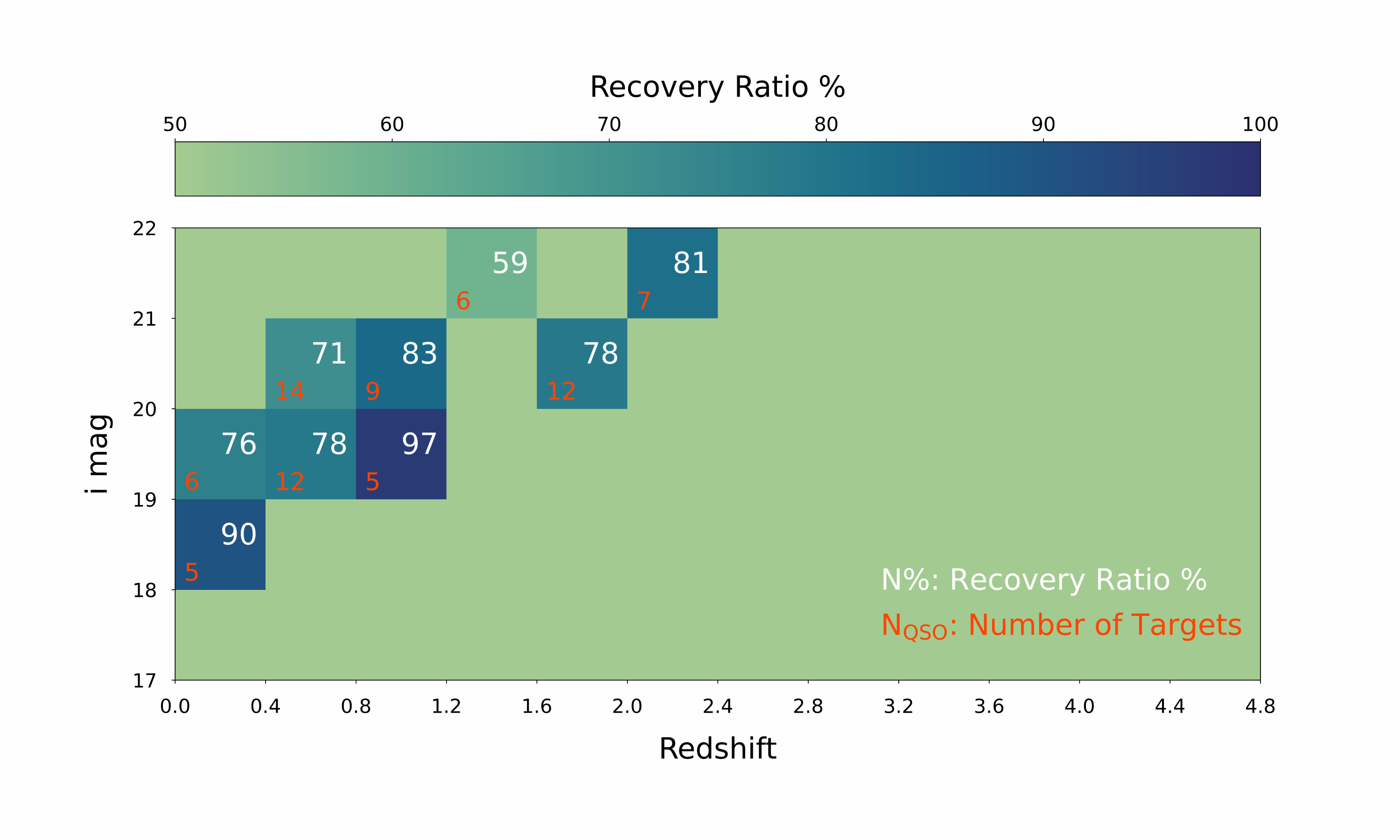}
 \caption{Overview of the lag recovery ratio binned by redshift and $i$-band magnitude. Here each bin reports the ratio of statistically significant lag measurements that are consistent to within 1$\sigma$ of the reported lag measurement in \citet{Shen2024}. Bins with fewer than five targets, mostly near boundary of redshift–magnitude parameter space, are excluded due to limited statistical significance. Overall, our light curve simulations (see Section \ref{sec:reduced cadence impact} for detail) shows that reducing the observation cadence by $\sim$ 40\% during the first year would result in slightly worse lag recovery ratios for $z<0.8$, where the \Hb\ emission line is primary line of interest for ground-based campaigns, and often exhibits shorter lags. The result of our simulations may be conservative since they are based on the gold sample in \citet{Shen2024}, which results in $\sim$ 100 in \Hb, \ion{Mg}{2}, and \ion{C}{4} emission lines.}
\label{fig:recovery}
\end{figure}

%\begin{figure*}
%\includegraphics[width=1.0\linewidth]{fig15_z_score_hb.pdf}
%\includegraphics[width=1.0\linewidth]{fig15_z_score_mg2.pdf}
%\includegraphics[width=1.0\linewidth]{fig15_z_score_c4.pdf}
% \caption{Same as Fig. 12, except that the panels on the right column show the distribution of the z-score among all the simulations over all of our targets, where the z-score is defined as $(\tau_{ori} - \tau_{sim})/\sigma_{ori}$. Here the long-lag and short-lag groups are separated. The dashed line show the $1\sigma$ errors of the original lag measurements. $\rm{N_{Long,|z|\leq1}, N_{Short,|z|\leq1}}$ show the ratios that $|z|<1$ of all significant results in the long-lag group and short-lag group.}
%\label{...}
%The quality of simulated lags for the gold sample of H$\beta$ (top), \ion{Mg}{2} (middle) and \ion{C}{4} (bottom). The panels in the left column show the changes of significant levels after epoch removal for each target. The dotted point represents the original significant level, while the cross represents the minimum significant level of 50 simulations for each target. $N_{sig}\%$ is the ratio of significant simulations. The panels in the middle column display the distribution of the consistency level for long-lag and short-lag groups in all significant simulations. Results between the dashed lines are consistent with the original lags in $1\sigma$ errors. $N_{Long,|\sigma|\leq1}\%, N_{Short,|\sigma|\leq1}\%$ are the ratios about simulations within $1 \sigma$ errors of all significant results in the long-lag group and short-lag group, respectively.
%\end{figure*}

\subsection{Impact of Reduced Cadence on Lag Success}\label{sec:reduced cadence impact}

We run all the simulated light curves through the same lag identification pipeline as \citet{Shen2024} and perform the following statistical analyses. We use two criteria to assess the difference between the original lag measurements and the simulated lag measurements for targets from the gold sample, and display the variation of simulated lags in Figure \ref{fig:sim_lag}. The two criteria are: 

\begin{itemize}
    \item The simulated lag is still detected with the statistically significant criteria defined in Section \ref{sec:timeseries}: (1) less than half of the lag posterior samples are removed by our alias removal process (as defined by the $f_{peak}$ parameter in Section~\ref{sec:timeseries}); (2) The maximum cross correlation coefficient $\rm{r_{max}}>0.4$ within $\pm 1 \sigma$ of the reported \texttt{PyROA} lag; (3) The \texttt{PyROA} lag is at least $2 \sigma$ different from zero lag, i.e., $\rm{|S/N|} > 2$, and for \texttt{PyROA}, it always means that the lag is positive at $\ge 2\sigma$ significance \citep{Shen2024};.
    
    \item The simulated lag is consistent with the original lag within $1\sigma$ uncertainties. To be more specific, we use consistency level $\tilde{\sigma}$ to quantitatively describe how well the original lag and the simulated lag are consistent within a given margin of error.

    If the simulated lag $\tau_{sim}$ is larger than the original lag $\tau_{ori}$, $\tilde{\sigma} = (\tau_{sim} - \tau_{ori})/(\sigma_{ori,upper}+\sigma_{sim,lower}) > 0$, where $\sigma_{ori,upper},\sigma_{sim,lower}$ is the upper uncertainties of original lag and lower uncertainties of simulated lags, respectively; on the contrary, if the simulated lag $\tau_{sim}$ is smaller than the original lag $\tau_{ori}$, $\tilde{\sigma} = (\tau_{sim} - \tau_{ori})/(\sigma_{ori,lower}+\sigma_{sim,upper}) < 0$, where $\sigma_{ori,lower},\sigma_{sim,upper}$ is the lower uncertainties of original lag and upper uncertainties of simulated lags, respectively. 
    
    %\item We determine the relative difference between the median of the 50 simulated lag $\tau_{sim}$ and the original lag $\tau_{ori}$ as $\delta = (\tau_{sim} - \tau_{ori})/\tau_{ori}$, and require the relative difference less than 34\%.
\end{itemize}

The lag selection criteria, specifically the $|\rm{S/N}|$, and $\rm{r_{max}}$ and their change, are shown in the top panels of Figure~\ref{fig:sim_lag}.
%The \textbf{change in} lag selection criteria\textbf{:} $|\rm{S/N}|$, and $\rm{r_{max}}$ are displayed on the \textbf{top} panels in Figure~\ref{fig:sim_lag}. 
Most of the lag posterior samples given by \texttt{PyROA} will still have a single, prominent peak in the lower-cadence simulations, so we do not show the changes for the lag selection criteria $f_{peak}$. The top panel of Figure~\ref{fig:sim_lag} displays the minimum statistical value across 50 simulations to capture the full range of lag selection criteria changes and resulting lag possibilities, potentially providing a better indication of the overall trend in a worst-case scenario. While the statistical values for $|\rm{S/N}|$ and $\rm{r_{max}}$ generally decreased in the lower-cadence simulations, most simulated lags still met the significance criteria. Figure~\ref{fig:sim_lag} highlights several targets with original $|\rm{S/N}|$ and $\rm{r_{max}}$
values near the selection threshold that resulted in insignificant lag measurements under reduced cadence conditions. The simulated lags that were initially considered significant but later became insignificant are primarily found in cases with lower overall light curve quality. This suggests that the reduction in measurement quality during the first season pushed these lags below the original selection criteria. Despite having the lowest median observed-frame lag (31 days), the \Hb\ sample still has the highest ratio of significant results. This may be attributed to the generally higher lag $|\rm{S/N}|$ and $\rm{r_{max}}$ values in the \Hb\ sample, indicating its robustness to epoch reduction. %\q{and jusstifying a potential reduction in the lag significance criteria.}

Our reduced-cadence simulations demonstrate that \Hb\ lags are most resilient to a reduction in number of epochs, with 94\% remaining significant after a 40\% epoch reduction in the first year of SDSS-RM. For \ion{Mg}{2} and \ion{C}{4}, we identified 88\% and 90\% of significant lags, respectively, using the criteria from \citet{Shen2024}.
Overall, our simulations demonstrate that we can recover $\sim$~90\% of significant results for all the three emission lines: \Hb, \ion{Mg}{2}, and \ion{C}{4}. \Hb\ lags exhibit the highest ratio of significant results, regardless of the original lag value. While targets with shorter original lags might be expected to be more affected by epoch reduction, we found no significant difference in the success rate for detecting long or short lags. %(the median original lags for \Hb, \ion{Mg}{2}, and \ion{C}{4} are 31 days, 155 days, and 140 days, respectively)

We display the consistency level $\tilde{\sigma}$ for significant results in the bottom row of Figure~\ref{fig:sim_lag}. As described above, the $\tilde{\sigma}$ can quantitatively describe how well the original lag and the simulated lag are consistent within a given margin of error. 
%For example, if the $|\tilde{\sigma}|< 1$, the original lag distribution and the simulated lag distribution are overlapping within $1\sigma$ errors. The positive (negative) value shows that the median of simulated lag distribution $\tau_{sim}$ is larger (smaller) than that of original lag distribution $\tau_{original}$. \yz{I changed the original text to this. I think it could have been written slightly more succinctly because we gave the equations of $\tilde{\sigma}$ above. You can decide which one is better.} 
%We also divided the sample into long and short-lag groups (separated by 31 days for \Hb, 155 days for \ion{Mg}{2}, and 140 days for \ion{C}{4} - the median original lag of each line). We found that the ratio of consistent results ($|\tilde{\sigma}| < 1$) for short-lag sample is slightly higher than than the long-lag sample for each emission line. However, when comparing the the ratios of consistent lag results for the three emission lines, we find that the simulation light curves of \ion{C}{4}, having a higher median original lag, resulting in the highest ratio of simulations that can still provide consistent results within $1\sigma$ errors, and the \Hb\ and \ion{Mg}{2} have lower ratios. 
We divided the sample into long- and short-lag groups based on their median original lags: 31 days for \Hb, 155 days for \ion{Mg}{2}, and 140 days for \ion{C}{4}. Short-lag targets exhibited a higher degree of consistency ($|\tilde{\sigma}| < 1$) compared to long-lag targets across all emission lines. The most pronounced difference was seen for \Hb, with short-lag targets achieving a $92\%^{+5\%}_{-4\%}$ consistency rate, surpassing the $79\% \pm 8\%$  of long-lag targets. \ion{Mg}{2} and \ion{C}{4} also exhibited similar trends, with short-lag targets outperforming long-lag targets by small margins. Short \ion{Mg}{2} lags showed a consistency rate of $87\%^{+7\%}_{-9\%}$, compared to $85\%^{+8\%}_{-7\%}$ for long \ion{Mg}{2} lags. Similarly, short \ion{C}{4} lags had a rate of $97\%\pm2\%$, while long \ion{C}{4} lags were at $95\%^{+4\%}_{-3\%}$.
%For each emission line, short-lag targets exhibited slightly higher consistency ($|\tilde{\sigma}| < 1$) than long-lag targets for all emission lines. For \Hb\ the consistency level for long lags are 79\% $\pm$ 1\%, while for short lags it is 92\% $\pm$ 1\%. For \ion{Mg}{2} the consistency level for long lags are 85\% $\pm$ 1\% and for short lags 87\% $\pm$ 1\%, and for \ion{C}{4} the long lags are consistent for 95\% $\pm$1\% and 97\% $\pm$ 1\%. 
Among the three emission lines, \ion{C}{4} (with the longest median original lag) demonstrated the highest overall ratio of simulations that maintained consistent results within 1$\sigma$ errors, while \Hb\ and \ion{Mg}{2} exhibited lower ratios. The short-lag sample may exhibit higher consistency due to its smaller absolute changes in lags and associated errors compared to the long-lag sample. Overall, if we require the simulated lags are still statistically significant and consistent with original lags in $\pm 1\sigma$, we can still recover 81\%, 76\%, and 86\% of all simulations for \Hb, \ion{Mg}{2}, and  for \ion{C}{4} respectively.%, which might have larger absolute difference of lags but also larger (absolute) lag uncertainties after cadence reduction.

To inform future cadence design, we examined the relationship between lag recovery ratio, $i$-band magnitude, and redshift, given that SDSS-RM targets are selected based on flux limits. Here, the lag recovery ratio is defined as the fraction of simulations in which the new lags obtained from simulations remain significant and consistent with the original lags within 1$\sigma$ (i.e., $|\tilde{\sigma}| < 1$) out of the total number of simulations.  Figure~\ref{fig:recovery} presents an overview of the lag recovery ratio binned by redshift (bin size = 0.4) and $i$-band magnitude (bin size = 1). Overall, we observed a trend of higher success ratios for brighter $i$-band magnitudes and lower redshifts. %This is expected because when redshift and i-band magnitude get smaller, the apparent brightness is higher and gives a larger lag if the target follows the Radius-Luminosity relation \citep{Bentz2013} for the AGNs. 
Several bins, particularly those near the boundary of the redshift-magnitude parameter space, contain fewer than five targets and may not provide statistically meaningful results. We therefore excluded these bins from Figure~\ref{fig:recovery}. Given our results based on seven years of observations from SDSS-RM, the limited number of significant lags detected at these redshifts and magnitudes, as also indicated by mock data in \citet{Shen2015a}, underscores the challenges of measuring significant lags at these cosmological distances and magnitudes. %Most bins with only 1-2 targets have very high recovery ratios, indicating that the number of detected targets is small at the redshifts and magnitudes of these bins, but their light curves are ideal for RM to provide reliable lags before and after the removal of epochs. Therefore, even though there are few targets in some bins, the recovery rates for those bins would still provide information about the efficiency of RM based on SDSS-RM data quality. Consequently, we would like to keep the bins with only a few targets and make a comparison with the results in \citet{Shen2015a}.

Our simulation results demonstrate a similar redshift and magnitude dependence for the recovery ratio as found in \citet{Shen2015a}, even when comparing our 7-year observations to their 6-month mock light curves (Figures 8 and 9 in \citealt{Shen2015a}).

%%%%%%%%%%%%%%%%%%%%%%%%%%%%%%%%%%%%%%%%%%%%%
%%%%%%%%%%%%%%%%%%%%%%%%%%%%%%%%%%%%%%%%%%%%%
\section{Summary and Implications for Future Industrial-Scale Surveys} \label{sec:discussion}
%%%%%%%%%%%%%%%%%%%%%%%%%%%%%%%%%%%%%%%%%%%%%
%%%%%%%%%%%%%%%%%%%%%%%%%%%%%%%%%%%%%%%%%%%%%
The Sloan Digital Sky Survey Reverberation Mapping (SDSS-RM) project has offered a valuable dataset for investigating factors that influence the successful measurement of reliable RM lags in quasars. Previous SDSS-RM studies \citep{Grier2017, Grier2019, Homayouni2020, Shen2024} have identified 172 high-confidence lag measurements (the gold sample). The gold lag success rate in SDSS-RM varies between 4\% and 20\% depending on the emission line type and survey length (see Table~\ref{tab:table1}). To better understand the gold-lag success rate, and analysis turnaround, we have analyzed the impact of intrinsic target properties, light-curve statistics, observation cadence, and the number of observing epochs on lag measurement. Our goal is to enhance the efficiency of lag analysis by identifying targets and light curves with a higher probability of success. This will facilitate a more strategic approach, allowing for efficient prioritization of targets and ultimately maximizing the scientific return of surveys that are similar to SDSS-RM.

For each individual lag study, we examined the correlation between target properties (luminosity, equivalent width of the targeted emission line, and redshift) and light curve characteristics (fractional RMS variability, SNR2, and Durbin-Watson statistics) with the success of measuring gold lags. Comparison of the target properties such as luminosity \lum\ (Figure~\ref{fig:luminosity}) and emission line equivalent width (Figure~\ref{fig:ew}) does not reveal a correlation with the gold-lag measurements. We observed a significantly higher success rate ($\gtrsim$20\%) in detecting gold lags when the \Hb\ line is positioned near the center of the observed-frame spectroscopic range. However, the impact of emission line position on detection is less strong for \ion{Mg}{2} due to redshift cuts and for \ion{C}{4} due to potentially longer lags exceeding our baseline coverage (Figure~\ref{fig:frac_var}).

As for the light curve characteristic, while fractional RMS variability shows no correlation with successful lag measurement, the empirical variability metric SNR2 exhibits a positive trend. The \Hb\ SNR2 in the gold sample of \citet{Grier2017} and \citet{Shen2024} is significantly higher for the gold lags than those of the parent sample. This is aligned with \citet{Shen2024}'s observations on the role of variability metrics as measured by SNR2 for general lag detection. Similarly, the gold lags measured by \citet{Grier2019} and \citet{Homayouni2020} maybe inconclusive for the impact of SNR2 as their initial samples were restricted to only have SNR2$> 20$.

The Durbin-Watson ($dw$) statistic is a reliable indicator of gold-lag measurements, particularly when applied to emission line light curves. While the $dw$ statistic for the continuum light curve shows a weaker correlation, analyzing the emission line $dw$ reveals a stronger trend. For \ion{Mg}{2}, the $dw$ statistic exhibits a weaker correlation with gold lag success compared to \Hb\ and \ion{C}{4}. This is likely due to \ion{Mg}{2}'s lower variability amplitude \citep{Sun2015}. Overall, our analysis suggests that the $dw$ statistic, especially when applied to emission line light curves, is a valuable tool for efficiently filtering targets and predicting the likelihood of obtaining reliable lag measurements, particularly in large datasets. Logistic regression further validates the \textit{dw} emission line as a strong predictor but also indicates a link with continuum luminosity (see Section\ref{sec:log_reg}). We caution, however, that the inverse correlation to source luminosity may be attributed to the longer time delays inherent in brighter sources (e.g., \citealt{Bentz2013}), potentially exceeding the observational baselines of current studies.

Additionally, we investigate the impact of reducing the SDSS-RM cadence on lag measurements based on the \textit{observed} quasar light curves of \citep{Shen2024}. We simulate multi-year light curves with 40\% reduction in cadence during the first year and compare with the existing lag measurements for the same target. While most statistically significant lags are preserved after cadence removal ($\sim$90\% or higher for all emission lines studied: \Hb, \ion{Mg}{2}, and \ion{C}{4}), a small percentage become insignificant. The  consistency between original and simulated lags (within a margin of error) is generally higher for shorter lags compared to longer lags. %However, when we consider the lag similarity, it is worse for shorter lags. This is likely due to the decreased temporal resolution, which can limit the ability to capture the rapid variations associated with shorter lags. 
Furthermore, we find that the recovery rate is slightly lower for fainter or higher redshift quasars (Figure~\ref{fig:recovery}). These results are consistent with earlier studies using shorter mock light curves \citet{Shen2015a}. Overall, our simulations indicate that a mixed cadence strategy, featuring a period of higher cadence followed by lower cadence as in SDSS-RM, provides only a marginal improvement in gold-lag measurement compared to a more uniform cadence, which can still recover 90\% of the lags.
%our simulations suggest that reducing the observing cadence in future industrial-scale RM surveys may be a viable option with minimal impact on lag measurement success rates.

\begin{acknowledgments}
The spectroscopic observations used in this study were obtained by the Sloan Digital Sky Survey (SDSS) from
2015$-$2020.  Funding for the Sloan Digital Sky Survey has been provided by the Alfred P. Sloan Foundation, the Heising-Simons Foundation, the National Science Foundation, and the Participating Institutions. SDSS acknowledges support and resources from the Center for High-Performance Computing at the University of Utah. SDSS telescopes are located at Apache Point Observatory, funded by the Astrophysical Research Consortium and operated by New Mexico State University, and at Las Campanas Observatory, operated by the Carnegie Institution for Science. The SDSS web site is www.sdss.org.

Y.H. was supported as an Eberly Research Fellow by the Eberly College of Science at the Pennsylvania State University. WNB thanks NSF grants AST-2106990 and AST-2407089. LCH was supported by the National Science Foundation of China (11991052, 12233001), the National Key R\&D Program of China (2022YFF0503401), and the China Manned Space Project (CMS-CSST-2021-A04, CMS-CSST-2021-A06).
\end{acknowledgments}

\bibliography{main.bib}
\end{document}